\documentclass[sn-basic]{sn-jnl}

\setcitestyle{aysep={},yysep={;}}

\usepackage{graphicx}%
\usepackage{multirow}%
\usepackage{amsmath,amssymb,amsfonts}%
\usepackage{amsthm}%
\usepackage{mathrsfs}%
\usepackage[title]{appendix}%
\usepackage{xcolor}%
\usepackage{textcomp}%
\usepackage{manyfoot}%
\usepackage{booktabs}%
\usepackage{algorithm}%
\usepackage{algorithmicx}%
\usepackage{algpseudocode}%
\usepackage{listings}%
\usepackage{ragged2e}
\usepackage{natbib}
\usepackage{arydshln}

\theoremstyle{thmstyleone}%
\theoremstyle{thmstyletwo}%

\theoremstyle{thmstylethree}%

\begin{document}

\title[Expert-elicitation method for non-parametric joint priors using normalizing flows]{Expert-elicitation method for non-parametric joint priors using normalizing flows}


\author*[1]{\fnm{Florence} \sur{Bockting}\orcid{https://orcid.org/0000-0003-0924-6413}}\email{florence.bockting@tu-dortmund.de}

\author[2]{\fnm{Stefan T.} \sur{Radev}\orcid{https://orcid.org/0000-0002-6702-9559}}\email{stefan.radev93@gmail.com} 

\author[1]{\fnm{Paul-Christian} \sur{Bürkner}\orcid{https://orcid.org/0000-0001-5765-8995}}\email{paul.buerkner@gmail.com}

\affil*[1]{\orgdiv{Department of Statistics}, \orgname{TU Dortmund University}, \orgaddress{\street{Vogelpothsweg 87}, \city{Dortmund}, \postcode{44227}, \state{North Rhine-Westphalia}, \country{Germany}}}

\affil[2]{\orgdiv{Cognitive Science Department}, \orgname{Rensselaer Polytechnic Institute}, \orgaddress{\street{110 Eighth Street}, \city{Troy}, \postcode{12180}, \state{NY}, \country{United States}}}


\abstract{
We propose an expert-elicitation method for learning non-parametric joint prior distributions using normalizing flows. Normalizing flows are a class of generative models that enable exact, single-step density evaluation and can capture complex density functions through specialized deep neural networks. Building on our previously introduced simulation-based framework, we adapt and extend the methodology to accommodate non-parametric joint priors. Our framework thus supports the development of elicitation methods for learning both parametric and non-parametric priors, as well as independent or joint priors for model parameters. To evaluate the performance of the proposed method, we perform four simulation studies and present an evaluation pipeline that incorporates diagnostics and additional evaluation tools to support decision-making at each stage of the elicitation process.
}




\maketitle
\section{Introduction}
The Bayesian paradigm offers the possibility to incorporate \emph{prior knowledge} into a \emph{statistical model} through the specification of \emph{prior distributions}.
This possibility is a central advantage of the Bayesian paradigm 
 \citep{mikkola2023prior}, yet it also presents one of its most challenging aspects \citep{simpson2017penalising,lgorzata2015sensitivity,van2006prior}.
In the following, we define prior knowledge as the expertise provided by a \emph{domain expert} --- an individual with extensive knowledge of a specific subject matter \citep{falconer2022methods}. This knowledge can be represented in various forms, but to integrate it into a Bayesian model, we need to translate it into a formal mathematical language that can be expressed as a prior distribution over the model parameters \citep{perepolkin2023quantile,o2019expert,martin2012eliciting, garthwaite2005statistical}.

A distinct field of research, commonly known as \emph{(expert) prior elicitation}, has developed around the question of how to gather expert knowledge and translate it into appropriate prior distributions. This area of study has a long history, dating back to the 1960s \citep{winkler1967assessment, kadane1980interactive, kadane1998experiences}, and continues to be an active area of research today \citep{stefan2022practical, mikkola2023prior, falconer2022methods}.
\citet{garthwaite2005statistical} identified four key stages in a \emph{prior elicitation process}: 
\begin{enumerate}
    \item \textbf{Setup stage}: In this stage, the problem is defined, an expert is selected, and the quantities to be elicited from the expert (in this paper referred to as \emph{target quantities}) are determined;
    \item \textbf{Elicitation stage}: Here, the target quantities are queried from the expert using specific elicitation techniques, resulting in what we call \emph{elicited statistics};
    \item \textbf{Fitting stage}: This involves fitting a (potentially joint) probability distribution based on the expert-elicited statistics;
    \item \textbf{Evaluation stage}: Finally, the adequacy of the fitted probability distribution is assessed in collaboration with the expert.
\end{enumerate}
In this context, \emph{elicitation methods} aim to provide a systematic and formal procedure for deriving prior distributions based on expert-elicited statistics. 
Early elicitation methods primarily tackled the problem by seeking analytical solutions, such as conjugate models or problem-specific transformation functions resulting in highly model-specific prior elicitation methods \cite[see][for a recent review]{mikkola2023prior}.
This model dependence is partly due to the use of direct elicitation techniques, which involve asking experts directly about model parameters \citep{stefan2022practical,falconer2022methods}. This approach is problematic not only due to its inherent model dependence but also because model parameters are often challenging for domain experts to interpret meaningfully. Consequently, the quality of expert input can be questionable. In response to these limitations, substantial efforts have been made to develop methods that enable more flexible and interpretable prior specification for domain experts.

A recently proposed group of elicitation methods uses advances in machine learning to automate the process of translating expert knowledge into prior distributions \citep{hartmann2020flexible,da2019prior,manderson2023translating,bockting2023simulation}. These methods address both key challenges: they allow expert knowledge to be expressed in terms of observable quantities and ensure that the elicited information remains interpretable for domain experts. Additionally, since the quantities elicited from the expert are not tied to specific model parameters, these methods are highly versatile and model-independent.
The underlying idea behind this approach is closely related to prior predictive checks, which are an integral part of the  \emph{Bayesian workflow} \citep{gelman2020bayesian,gabry2019visualization}. The key concept is the \emph{prior predictive distribution} (PPD), defined as $p(y)=\int p(y \mid \theta) p(\theta) d\theta$ with likelihood $p(y\mid \theta)$ and prior $p(\theta)$. The PPD establishes a formal relationship between the prior distributions of the model parameters and the model predictions, $p(y)$. In this way, the PPD provides a means to link expert knowledge, expressed in terms of observable quantities, to the latent model parameters. 
Since the PPD is often not analytically tractable, it is typically approximated using Monte Carlo integration \citep{mikkola2023prior} which involves first sampling $\theta' \sim p(\theta)$ and then $y'\sim p(y \mid \theta')$. The resulting model predictions, $y'$, are then compared to expert knowledge on the outcome variable, $y^*$, using an appropriate discrepancy loss function. The goal is to learn prior distributions, $p(\theta)$, that produce model predictions consistent with expert knowledge by means of minimizing the discrepancy loss \citep{garthwaite2005statistical, gelman2017prior,simpson2017penalising,betancourt2020towards}.

The methods introduced so far that use this \emph{simulation-based} approach focus on learning independent, parametric priors $p(\theta \mid \lambda)$, parameterized by a set of prior hyperparameters $\lambda$ \citep{hartmann2020flexible,da2019prior,manderson2023translating}. While assuming \emph{independent} priors for model parameters can be reasonable in some cases due to theoretical considerations or model constraints, this approach may not be sufficient for more complex or high-dimensional problems \citep{gelman2017prior,gelman2020bayesian,simpson2017penalising}. In such cases, it becomes important to account for the \emph{joint distribution} of model parameters to capture dependencies and interactions that reflect the underlying structure more accurately.
Similarly, the use of \emph{parametric} prior distribution families can be well-justified in some cases, but may lead to model misspecification in others, prompting the need for \emph{non-parametric} approaches. Elicitation methods that focus on non-parametric priors are less studied, but examples include the use of Gaussian processes \citep{oakley2007uncertainty} and quantile-parameterized distributions \citep{perepolkin2023quantile,perepolkin2024hybrid}.

Building in particular on the work of \citet{hartmann2020flexible,da2019prior,manderson2023translating}, we previously introduced a simulation-based framework with a highly modular structure that closely aligns with the elicitation process described by \citet{garthwaite2005statistical}. The advantage of this modular structure is its flexibility, allowing for easy adaptation to different types of methods. In our previous work \citep{bockting2023simulation}, we demonstrated how to use this framework to learn parametric prior distributions for the model parameters. In this paper, we show how the same simulation-based framework can be used to learn flexible (i.e., non-parametric) joint priors for the model parameters with only minor adjustments to the workflow. In both approaches, we employ mini-batch stochastic gradient descent as the optimization method. 

\paragraph{Main Contributions}
Our motivation is two-fold:
First, we aim to promote the development of a unifying, highly modular framework that supports a wide range of prior elicitation methods. We believe that such a framework would benefit both users and developers of prior elicitation methods by providing structure and clarity in a field currently populated with numerous coexisting methods, which can make comparisons challenging.
Second, we want to demonstrate how our simulation-based framework, introduced in \citet{bockting2023simulation}, can serve as a flexible approach for learning either parametric or non-parametric, as well as independent or joint, priors with only minor adjustments. While our previous work focused on learning parametric, independent priors, the present work emphasizes learning non-parametric, joint priors.

\section{Methodology} \label{sec:methodology}
In this paper, we propose an elicitation method that extends our recently introduced simulation-based framework \citep{bockting2023simulation} to support the learning of non-parametric joint priors. The overall workflow remains unchanged; the modifications primarily concern how the prior distributions are specified and learned. We will first provide a brief overview of the workflow to establish the background, followed by a more detailed discussion of the modification introduced to the original framework, which is the focus of the current paper.

\begin{figure}[ht]
    \centering
    \includegraphics[width=1.\linewidth]{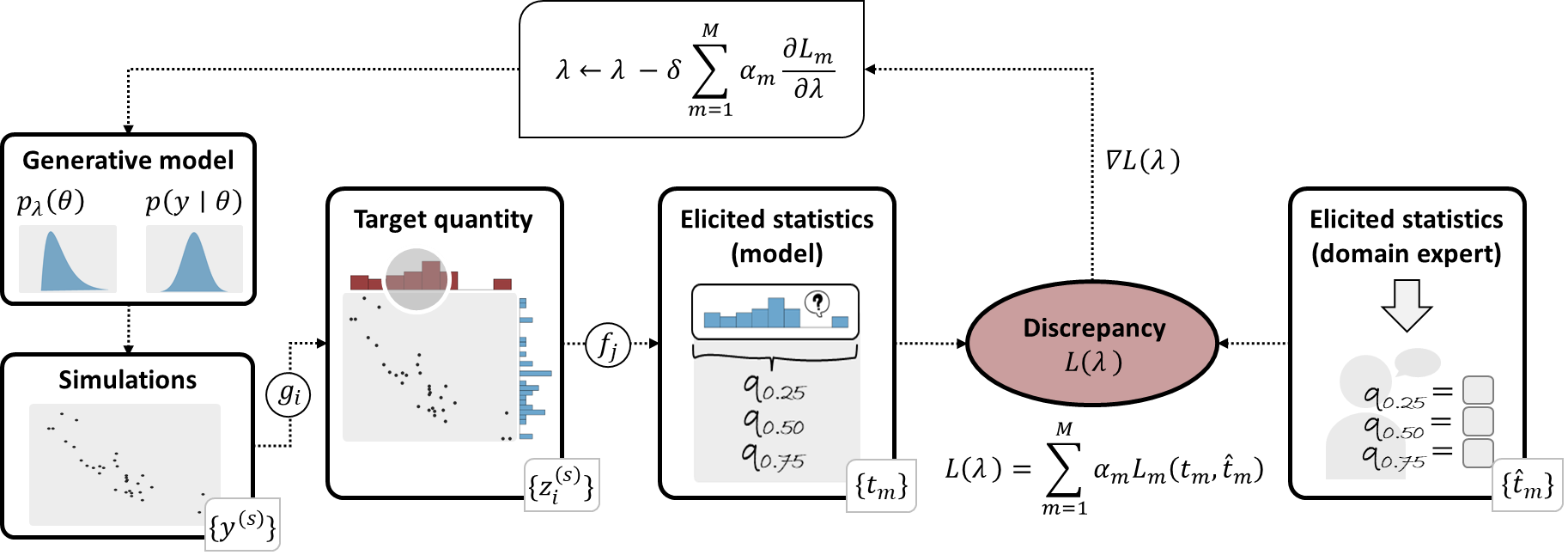}
    \caption{\emph{Graphical illustration of our simulation-based prior-elicitation framework.} 
    The process begins by identifying target quantities to be elicited from the domain expert and selecting appropriate elicitation techniques, which result in (expert-)elicited statistics. Next, predictions are simulated from the generative model by sampling from the prior $p_\lambda(\theta)$ and computing the corresponding model-implied target quantities and elicited statistics. The consistency between model-and expert-elicited statistics is assessed using a loss function $L_m$, where $\alpha_m$ is the weight of the $m^\text{th}$ loss component. Hyperparameters $\lambda$ that define the learned prior are adjusted based on this evaluation to reduce the loss and align model predictions more closely with expert knowledge. This iterative process continues until a prior is found that minimizes the discrepancy between model and expert-elicited statistics. In the updating rule, $\delta$ refers to the step size.}
    \label{fig:concept}
\end{figure}

The general workflow of the framework closely resembles the approach of \emph{prior predictive checks} \citep{gelman2020bayesian}: We simulate from the joint model $p(\theta, y)$ and assess how well the resulting prior predictions align with the expert's expectations. If there is a discrepancy between the expert's expectations and the model simulations, the prior specification needs to be adjusted accordingly.
Figure~\ref{fig:concept} provides a graphical representation of the framework and Table~\ref{tab:symbols} offers a summary of the symbols and notation used in the following sections.
The general workflow of our framework can be summarized as follows:
\begin{enumerate}
    \item \emph{Define the generative model}: Define the generative model including dimensionality and parameterization of prior distribution(s). (Setup stage; Section~\ref{subsec: prior-specification})
    \item \emph{Identify variables and elicitation techniques for querying expert knowledge}: Select the set of variables to be elicited from the domain expert (target quantities) and determine which elicitation techniques to use for querying the selected variables from the expert (elicited statistics). (Setup stage; Section~\ref{subsec: target-quantities-elicitation-techniques})
    \item \emph{Elicit statistics from expert and simulate corresponding predictions from the generative model}:
    Sample from the generative model and perform all necessary computational steps to generate model predictions (model-elicited statistics) corresponding to the set of expert-elicited statistics. (Elicitation stage; Section~\ref{subsec: model-implied-expert-elicited-statistics})
    \item \emph{Evaluate consistency between expert knowledge and model predictions}: Evaluate the discrepancy between the model- and expert-elicited statistics via a multi-objective loss function. (Fitting stage; Section~\ref{subsec: discrepancy-prior-learning})
    \item \emph{Adjust prior to align model predictions more closely with expert knowledge}: Use mini-batch stochastic gradient descent to adjust the prior so as to reduce the loss. (Fitting stage; Section~\ref{subsec: discrepancy-prior-learning})
    \item \emph{Find prior that minimizes the discrepancy between expert knowledge and model predictions}: Repeat steps 2 to 5 iteratively until a prior is found that minimizes the discrepancy between the model and expert-elicited statistics. (Fitting stage; Section~\ref{subsec: discrepancy-prior-learning})
    \item \emph{Evaluate the learned prior distributions}: Run the learning algorithm (steps 2 to 6) multiple times to obtain a set of prior distributions that can equally well represent the expert data. Select a plausible prior distribution in consultation with the domain expert or apply model averaging techniques. (Evaluation stage; Section~\ref{subsec: evaluation-results})
\end{enumerate}
The following sections discuss each step in greater detail.

\begin{table}[h!]
\begin{tabular}{p{0.4\textwidth} p{0.5\textwidth}}
\hline
\emph{Symbol}  &  \emph{Description}\\
\hline
\multicolumn{2}{l}{Simulation-based prior elicitation}\\
\hline
$y_n$ & data $n = 1, \ldots, N$\\
$\theta_k$ & model parameter $k=1,\ldots,K$ \\
$\lambda$ & model hyperparameter \\
$p(y \mid \theta)$  & likelihood \\
$p(\theta \mid \lambda)$ & prior parameterized by $\lambda$ \\
$p(y)$ & prior predictive distribution \\
$z_i = c_i(\lambda)$ & simulated model target quantity defined by function $c_i$ \\
$\hat z_i$ & expert representation of the target quantity \\
$t_m = f_j(z_i)$ & model-elicited statistic defined by elicitation technique $f_j$\\
$\hat t_m$ & expert-elicited statistic \\
$L_m(t_m(\lambda),\hat t_m)$ & loss component \\
$L(\lambda)=\sum_{m=1}^M\alpha_m L_m(t_m(\lambda),\hat t_m)$ & total loss as weighted sum of $L_m$ with weights $\alpha_m$\\
$p_{\lambda_1,\ldots,\lambda_R}(\theta)=\sum_{r=1}^R w_r \cdot p_{\lambda_r}(\theta)$ & averaged prior across $R$ replications and weighted by $w_r$ with $r=1,\ldots,R$\\
$\Delta_r(L) = L(\lambda_r)-L(\lambda_{\min})$ & difference between total loss of replication $r$ and replication with minimum total loss \\
$w_r = \frac{\exp\{-\gamma \Delta_r(L)\}}{\sum_{v=1}^V\exp\{-\gamma \Delta_v(L)\}}$ & weights used for model averaging with $\gamma$ being a scaling factor; in all simulation studies $\gamma=1.$\\
\hline
\multicolumn{2}{l}{Normalizing flow} \\
\hline
$u\sim p_U(u)$ & samples from base distribution \\
$g_\lambda(\cdot)$ & generator function \\
$u = g_\lambda(\theta)$ & normalizing direction \\
$\theta = g^{-1}_\lambda(u)$ & generative direction \\
$g_\lambda = g_{\lambda_H} \circ \cdots \circ g_{\lambda_1}$ & composition of $H$ coupling layers\\
\hline
\end{tabular}
\caption{Notation and symbols used in this paper.}
\label{tab:symbols}
\end{table}

\subsection{Prior Specification \& Learning}
\label{subsec: prior-specification}
In our previous work \citep{bockting2023simulation}, we introduced the simulation-based framework along with a method for learning a set of \emph{independent parametric} prior distributions, $p(\theta_k \mid \lambda_k)$, for the model parameters $\theta_k$ with $k=1,\ldots,K$. 
In this paper, we extend our framework by introducing a method to learn a \emph{joint non-parametric} prior distribution $p_\lambda(\theta) = p_\lambda(\theta_1,\ldots,\theta_K)$ over all model parameters $\theta$. 
In both the parametric and non-parametric approaches, the objective of the optimization process is to learn the hyperparameters $\lambda$ that minimizes the discrepancy between the model simulations and the expert expectations.
However, the interpretation of these hyperparameters differs between approaches, which we emphasize by using a different notation for the prior distributions.
In the non-parameteric approach, we employ \emph{normalizing flows} (NFs) to induce a flexible family of prior distributions which entail specialized deep neural networks with trainable parameters $\lambda$ \citep{kobyzev2020normalizing}.
Thus, in the non-parametric approach, $\lambda$ represents the weights of the deep neural networks within the NFs, whereas in the parametric approach, $\lambda$ denotes the parameters of the associated prior distribution families (i.e., the prior hyperparameters). A more detailed introduction to NFs will be provided in the upcoming section. 

In summary, the main difference between the parametric approach (proposed by \cite{bockting2023simulation}) and the non-parametric approach (focus of the current paper) is that, in the former, the user must specify a parametric prior distribution family for each model parameter, and the corresponding prior hyperparameters are learned via stochastic optimization. In the new extension, the user no longer needs to specify a parametric prior for the model parameters. Instead, a complex non-parametric prior distribution is assumed. This non-parametric joint prior is represented and learned using deep neural networks (specifically, NFs), with the weights learned via stochastic optimization. Table~\ref{tab:comparison-deep-parametric-prior} summarizes the main steps in the elicitation process and highlights where the extension modifies the existing workflow of our simulation-based framework. Furthermore, we provide in this paper additional diagnostics to support users in interpreting the simulation results.
\begin{table}[ht]
    \centering
    \begin{tabular}{p{0.2\textwidth} p{0.33\textwidth} p{0.32\textwidth}}
    \hline
     \emph{Elicitation process} & \emph{parametric-prior approach} & \emph{deep-prior approach} \\
     \hline
     Setup & & \\
     \quad \emph{Model} & specify generative model & \% \\
     \vspace{0.1cm}\\
     \quad \emph{Priors} & specify \textbf{parametric prior} distribution families & assume \textbf{non-parametric  joint prior} (specified via NFs) \\
     \vspace{0.1cm}\\
     \quad \emph{Data} & specify target quantities and elicited statistics & \% \\
     & (\textbf{less expert input} needed due to stronger regularized priors) & (\textbf{more expert input} required (e.g., dependency information) due to flexible prior) \\
     \vspace{0.1cm}\\
    \emph{Elicitation} & specify elicitation techniques used to query the domain expert & \% \\
    \vspace{0.1cm}\\
    \emph{Fitting} & apply mini-batch stochastic gradient descent & \% \\
    \vspace{0.1cm}\\
    \emph{Evaluation} & provide diagnostics, visualizations, etc. to evaluate simulation results & \% \\
    \hline
    \end{tabular}
    \caption{Comparison between the deep-prior and parametric-prior approach on a conceptual level. Note: The symbol `\%' refers to the same content as provided in parametric-prior approach. NFs is an abbreviation for Normalizing Flows (see text for details).}
    \label{tab:comparison-deep-parametric-prior}
\end{table}

\paragraph{Normalizing Flows}\label{subsec:affine-coupling-flows}
NFs transform a simple probability distribution, known as the \emph{base distribution} $p(u)$, into a more complex \emph{target distribution}, which, in our case, is the joint prior $p_\lambda(\theta)$. This transformation is achieved through a series of invertible and differentiable mappings $g_{\lambda}$, parameterized by $\lambda$ \citep{kobyzev2020normalizing}. 
The composition of invertible functions $g_\lambda$ yields an explicit form of the density $p_\lambda(\theta)$ via the change of variables formula
\begin{equation*}
p_\lambda(\theta) = p(u = g_\lambda(\theta)) \mid \det g_\lambda'(\theta) \mid,
\end{equation*} 
where $g_\lambda'(\theta) = \frac{\partial}{\partial\theta}g_\lambda(\theta)$ is the Jacobian matrix of $g_\lambda$ at $\theta$ and $\mid \det g_\lambda'(\theta) \mid$ is its absolute determinant. This formula allows us to easily evaluate the prior density at arbitrary points of interest $\theta$ \citep{dinh2016density}. Obtaining samples from the joint prior can be achieved by first sampling from the base distribution, $p(u)$, and then applying the inverse $g^{-1}_\lambda(u)$ to obtain samples from $p_\lambda(\theta)$:
\begin{equation*}
    \theta = g^{-1}_\lambda(u) \sim p_\lambda(\theta) \quad \text{for} \quad u \sim p(u).
\end{equation*} 
Typically, $p(u)$ is chosen to be a simple density, such as a standard Gaussian or a uniform distribution \citep{kobyzev2020normalizing}. 

For NFs to be practical, the mapping $g_\lambda$ should be easily invertible, computationally efficient, and sufficiently expressive to model any target distribution of interest \citep{kobyzev2020normalizing}. 
Several types of NFs that meet these requirements have been proposed, including affine \citep{dinh2014nice}  and spline coupling flows \citep{durkan2019neural}.
A detailed discussion and introduction to coupling flows can be found in various sources \citep{draxler2024universality, dinh2014nice, dinh2016density, radev2020bayesflow, kobyzev2020normalizing, papamakarios2021normalizing,bond2021deep}.
Due to their simple architecture, affine coupling flows are easy to invert and computationally efficient, although they have restricted expressiveness \citep{bond2021deep}. However, the expressiveness of a coupling flow can be increased by stacking multiple affine coupling layers, $g_\lambda = g_{\lambda_H} \circ \cdots \circ g_{\lambda_1}$ with $h=1,\ldots, H$ coupling blocks \citep{dinh2016density}. In fact, \citet{draxler2024universality} prove that affine coupling flows are distributional universal approximators, capable of representing any target distribution despite their seemingly restrictive architecture. In line with these considerations and the strong empirical performance of affine coupling flows, we focused on this type of normalizing flows in our simulation studies (see Section~\ref{sec:simulation-studies}), using the implementation provided by BayesFlow \citep{radev2023bayesflow}.

Finally, it is worth noting that, since our focus is solely on the mapping from the base distribution to the target distribution (i.e., the generative direction of NFs), virtually any other generative model family could be used to learn how to generate samples from a joint prior. Examples of common generative architecture alternatives include diffusion models \citep{cao2024survey} and flow matching \citep{lipman2022flow}. However, we used NFs here because they perform well in practice and are computationally efficient. 
In particular, they do not require solving differential equations during sampling \citep{draxler2024universality}.

\subsection{Target Quantities \& Elicitation Techniques}
\label{subsec: target-quantities-elicitation-techniques}
A key task in the setup stage of the elicitation process is the selection of target quantities and elicitation techniques to effectively gather the expert information. Two main criteria should guide this selection: \emph{interpretability} and \emph{informativeness} \citep{mikkola2023prior,crowder1992bayesian,da2019prior,garthwaite2005statistical}. 
\paragraph{Selecting Target Quantities}
\emph{Interpretability} refers to the selection of target quantities that allows domain experts to meaningfully express their knowledge based on their experience and expertise \citep{garthwaite2013prior}. Target quantities on the same scale as the outcome variable (cf., \emph{observable} quantities) are considered to be highly interpretable \citep{da2019prior,manderson2023translating}. In contrast, target quantities that refer to model parameters are more challenging to interpret, especially when the outcome variable is not on the same scale as the model parameters \citep{kadane1998experiences,kadane1980interactive}. 
Therefore, it is sometimes advocated asking experts exclusively about observable quantities \citep{kadane1998experiences}. We share the view of \citet{denham2007geographically} that the appropriateness of either approach depends on the specific problem being addressed and the type of expert providing the information. 

Guided by this idea, we define a target quantity within our framework in the most general sense as a function of the trainable hyperparameters of the prior: $z_i = c_i(\lambda)$ for $i=1,\ldots,I$. Typically, multiple target quantities ($I$ in total) are elicited from the user to inform different aspects of the data-generating process.
For instance, if we want to query the expert regarding the outcome variable (i.e., the observable space), one of the target quantities can be simply defined as $z_i=y$. This is obtained by defining the function $c_i$ as a query to the prior predictive distribution $p(y) = \int p(y \mid \theta) p_\lambda(\theta) d \theta$, where the likelihood is marginalized over the prior. Note that this expression is not just a constant as it refers to the prior predictive distribution and should not be confused with the marginal likelihood of observed data (see for this point also \citet{hartmann2020flexible}).
As another example, expert knowledge may be elicited about the parameter space. In this case, the target quantity is defined as $z_i=\theta_k$, which is obtained by defining $c_i$ as a simple projection onto $\theta_k$. 
In this way, we can also accommodate any other target quantity that can be derived from the model parameters or the data \citep[e.g., $R^2$; the proportion of variance explained by the model;][]{gelman2019r}.

\paragraph{Selecting Elicitation Techniques}
Once the set of target quantities has been selected, we need to determine how to elicit this information from the expert, which requires choosing an appropriate \emph{elicitation technique}. While experts can be asked to provide the full distribution of a target quantity (e.g., in the form of a histogram \citep{johnson2010methods}), another common technique is to focus on specific summary statistics \citep{morris2014web}. Research on prior elicitation suggests that experts can reasonably be asked to describe a target quantity using proportions, the mode, the median, and generally quantiles \citep{garthwaite2005statistical,stefan2022practical,o2006uncertain}, with  quantile-based elicitation techniques being particularly recommended \citep{kadane1998experiences}. In contrast, asking about the mean is less advisable when the distribution is skewed \citep{peterson1964mode}, and similarly, querying variances is in general discouraged \citep{garthwaite2005statistical, kadane1998experiences}. Furthermore, assessing dependencies between variables is challenging, making the elicitation of joint priors especially difficult \citep{garthwaite2005statistical}. Elicitation of correlations in the context of joint priors has been studied among others by \citet{gokhale1982assessment, clemen2000assessing}, and \citet{dickey1985bayesian} as reviewed by \citet{mikkola2023prior}.

In our elicitation method, we represent an elicitation technique as a function $f_j$ for $j=1,\ldots,J$, of a target quantity $z_i$. The resulting set of \emph{elicited} target quantities is referred to as \emph{elicited statistics}, ~$\{t_m(\lambda)\}$, where $t_m(\lambda) = f_j(z_i)$ and $m$ refers to the corresponding $i \times j$ combination. In the following, we sometimes omit the explicit dependence on $\lambda$ and simply write $t_m$ instead of $t_m(\lambda)$. Note also that, depending on the elicitation technique employed, $t_m$ may accept either a single value or a set of values as input. 
To illustrate this point more clearly, consider the case in which we elicit expert knowledge about the predictive distribution conditional on a category of a categorical predictor, i.e., $z = y\mid \text{cat}$. To elicit this information, we query the expert regarding the $25\%, 50\%$, and $75\%$ quantiles of this predictive distribution. Thus, we have $t = \{Q_{25\%}(y\mid \text{cat}), \, Q_{50\%}(y\mid \text{cat}), \, Q_{75\%}(y\mid \text{cat})\}$.

\paragraph{Sensitivity Analysis}
In addition to interpretability, the second criterion for selecting target quantities and elicitation techniques is \emph{informativeness}, which refers to the relevance of the elicited statistics for learning the prior distributions.
Evaluating this criterion is nontrivial because the relationship between the elicited statistics and the model (hyper-)parameters is complex and often analytically intractable. 
One computational approach to asses informativeness is to conduct a \emph{sensitivity analysis} \citep{depaoli2020importance}.
In this approach, we systematically vary one aspect of the prior distribution at a time while keeping all other aspects constant. For each specific change in the prior, we run the generative model in forward mode by sampling from the prior, then from the likelihood, and subsequently computing the target quantities and elicited statistics. Sensitivity analysis enables us to assess how changes in one aspect of the prior distribution impact each elicited statistic. If we observe no variation in a particular elicited statistic for a given prior variation, this statistic does not provide any useful information for the learning algorithm to determine the corresponding aspect of the prior. This approach can help inform the selection of elicited statistics and identify which aspects of the prior are most likely to remain unidentifiable given the current set of expert information.

\subsection{Model-implied \& Expert-elicited Statistics}
\label{subsec: model-implied-expert-elicited-statistics}
After determining the set of target quantities and corresponding elicitation techniques, we can proceed to gather information from the expert through specific \emph{prior elicitation protocols} \citep{gosling2018shelf,cooke1991experts,european2014guidance}. This process may be preceded by a training phase to familiarize the expert with the elicitation tasks \citep{stefan2022practical}. In prior elicitation research, an entire subfield has focused on determining how to conduct optimal expert interviews, taking into account factors such as cognitive biases and heuristics. While it is beyond the scope of this paper to review this literature, several comprehensive reviews are available \cite[e.g.,][]{mikkola2023prior, stefan2022practical, falconer2022methods}. At this point, we will skip the actual interrogation process and assume that the expert information has been successfully gathered, providing us with a set of expert-elicited statistics, individually denoted by $\hat t_m$ and together as $\{\hat t_m\}$. 

Given the set of expert-elicited statistics, $\{\hat t_m\}$, we can evaluate the discrepancy between the expert expectations and the model-implied statistics, $\{t_m\}$. The latter are computed by simulating from the model. This proceeds as follows:
We start with an arbitrary initial prior distribution from which we sample model parameters $\theta^{(s)} \sim p_{\lambda_0}(\theta)$, where $\lambda_0$ represent the initial hyperparameter values that define the prior distribution. The superscript $s$ denotes the $s^\text{th}$ sample out of a total of $S$ samples. Subsequently, we perform all necessary computational steps to derive samples from the target quantities $z_i^{(s)}$. Finally, we apply the corresponding elicitation technique to obtain the model-implied statistics $t_m = f_j(\{ z_i^{(s)} \})$ from the simulations.


\subsection{Discrepancy Evaluation \& Training}
\label{subsec: discrepancy-prior-learning}
To evaluate the discrepancy between the model- and expert-elicited statistics, an appropriate loss function is used for each statistic, $L_m(t_m(\lambda), \hat t_m)$. Different loss functions may be preferable depending on the type of the respective elicited statistic.  
Since the discrepancy is computed for each statistic within the set of elicited statistics, the total loss comprises multiple \emph{loss components}, which are combined into a single total loss using a weighted sum: 
\begin{equation}
L(\lambda) = \sum_{m=1}^M\alpha_m L_m(t_m(\lambda), \hat t_m),
\end{equation}
where $\alpha_m$ denote the weights. 
The choice of weights $\alpha_m$ can be guided by different considerations: One approach involves manually adjusting the weights, guided either by the expert's assessment of the importance of individual loss components or by ensuring that the effects of different magnitudes, arising from varying scales, are balanced, such that no single loss component dominates the others \citep{wang2011multi}. Alternatively, automated weighting methods, known as loss-balancing methods, can be used to optimally balance the contribution of each individual term to the total gradient \citep{liu2019end,crawshaw2020multi, bischof2021multi}.

After computing the total loss, $L(\lambda)$, the gradients with respect to the hyperparameters $\lambda$ are calculated. These gradients are then used to adjust the hyperparameters in the opposite direction: 
\begin{equation}
\lambda^{'} \leftarrow \lambda - \delta \sum^M \alpha_m \frac{\partial L_m}{\partial \lambda},
\end{equation}
with $\delta$ being the step size \citep{goodfellow2016deep}. We employ mini-batch stochastic gradient descent (SGD) with automatic differentiation, facilitated by the (explicit or implicit) reparameterization trick \citep{kingma2014auto,figurnov2018implicit}.
With the adjusted hyperparameters $\lambda'$, new samples from the prior $p_{\lambda'}(\theta)$ can be generated and the updated model-implied statistics $t_m(\lambda')$ computed. We then reassess the discrepancy between the model-implied and expert-elicited statistics and continue to adjust the hyperparameters as needed. This iterative process continues until a convergence criterion is met or a stopping rule is applied. 

\subsection{Evaluation of Training Results}\label{subsec: evaluation-results}
\paragraph{Convergence Checks}
Training is considered successful if the loss in minimized and no further learning occurs.
With a proper discrepancy measure as loss function, it is guaranteed that the total loss (and the individual loss components) approach zero as learning progresses.
The learning progress can be tracked by examining the loss behavior across epochs, where a decreasing trend is expected until the loss stabilizes at a value close to zero.
Further insights into the learning progress can be obtained by tracking the convergence of additional quantities of interest beside the loss, for instance the mean and standard deviation of the marginal priors. 

A further method to assist in the examination of convergence is to compute the slope of a linear regression fitted to the loss values over the last $m$ epochs, whereby $m$ should not be taken too large such that the linearity assumption holds. Typically, the overall loss function exhibits exponential decay, but once the loss stabilizes, a linear trend is a good approximation. Ideally, a slope of zero would indicate perfect convergence, though some variation is expected in practice. To evaluate whether the slope remains acceptable, the training algorithm can be run repeatedly with random seeds and the absolute slopes across all replications can be compared.
For training runs with the highest final slopes, we can perform a visual inspection of the trajectories of key quantities across epochs, such as the total loss, loss components, and prior means and standard deviations. If these runs show successful convergence, we can reasonably conclude that other runs (with smaller final slopes) have also converged.

\paragraph{Non-uniqueness of Learned Priors}
Once the elicited statistics have been learned accurately, we can examine the corresponding learned prior distribution. As there are often multiple optimal values $\lambda^*$ that align closely with the same set of elicited statistics, multiple replications will yield different prior distributions \citep{da2019prior,manderson2023translating,stefan2022practical}. This lack of uniqueness is anticipated, given the limited information provided by the set of elicited statistics compared to the complexity of the generative model \citep{manderson2023translating}. 
Furthermore, when target quantities refer solely to the observable space, we only gain access to the model's overall uncertainty with limited ability to differentiate between aleatoric and epistemic uncertainty \citep{nemani2023uncertainty,perepolkin2024hybrid}. Aleatoric uncertainty, by definition, is irreducible; however, several methods exist to reduce epistemic uncertainty. These include (i) selecting more informative target quantities, (ii) eliciting more detailed expert information (e.g., increasing the number of quantiles in quantile-based elicitation), (iii) adding theory-informed regularization terms to constrain the search space \citep[as illustrated in][]{manderson2023translating}, (iv) improving the initialization method to achieve better starting points, and (v) employing more advanced optimization algorithms when dealing with a large number of model parameters \citep{nemani2023uncertainty}.

\paragraph{Model Averaging}
Given the (potential) non-uniqueness of the learned prior distributions, it is advisable to run the training multiple times on the same expert data but with different random seeds. This approach helps provide insight into the variability among the learned prior distributions. From the resulting set of candidate priors $p_{\lambda_r}(\theta)$ for $r=1,\ldots,R$, we can, in consultation with the domain expert, exclude prior distributions that do not meet the expert's expectations. For the remaining set of prior distributions, which may be considered practically equivalent from the expert’s perspective, either a single prior distribution can be chosen, or alternatively, model averaging techniques can be applied. In the case of model averaging, the combined prior would take the form:
\begin{equation}\label{eq: model-averaging}
    p_{\lambda_1,\ldots,\lambda_R}(\theta) = \sum_{r=1}^R w_r \cdot p_{\lambda_r}(\theta)
\end{equation}
where the weights $w_r$ sum to one. The weights $w_r$ can either be assigned by the user, based on theoretical assumptions, or selected according to one of several proposed weighting schemes \citep{claeskens2008model}. Here, we apply an approach suggested by \citet{buckland1997model} for the general purpose of model averaging and selection, where the weights $w_r$ are defined as
\begin{equation*}
    w_r = \frac{\exp\{-\gamma I_r\}}{\sum_{v=1}^V \exp\{-\gamma I_v\}}.
\end{equation*}
In the above expression, $\gamma$ is a scaling factor, and $I$ represents the performance criterion by which the models are weighted. This definition ensures that models with identical performance receive equal weight \citep{buckland1997model}. In our setting, it is sensible to use the total loss $L(\lambda_r)$ of each replication $r$ as the performance criterion. Consequently, the ``best'' model is the one with the smallest total loss, $L(\lambda_{\min})$ \citep{bissiri2016general}. This yields the following definition of weights
\begin{equation}\label{eq: weights}
    w_r = \frac{\exp\{-\gamma \Delta_r(L)\}}{\sum_{v=1}^V \exp\{-\gamma \Delta_v(L)\}}.
\end{equation}
where $\Delta_r(L) = L(\lambda_r) - L(\lambda_{\min})$. The difference is used primarily for computational reasons to ensure numerical stability \citep{claeskens2008model}. The formulation in Eq.~(\ref{eq: weights}) can be interpreted as the probability of replication $r$ representing the best prior, given the collection of learned prior distributions \citep{wagenmakers2004aic}. 
The scaling factor $\gamma$ controls the influence of each replication’s relative performance $\Delta_r(L)$. When $\gamma>1$, discrepancies in model performance are amplified, with better models receiving higher weights and worse models receiving lower weights. As $\gamma \rightarrow \infty$, this weighting would approach a one-hot vector, effectively performing model selection. Conversely, when $\gamma<1$, the performance differences between models are smoothed out, reducing the influence of performance discrepancies.

\paragraph{Simulating From an ``Oracle''}
The simulation-based framework enables us to evaluate the degree of non-identifiability and determine if further regularization is required, even before querying any data from the expert.
An effective approach involves using an \emph{oracle}, where a pre-defined prior distribution serves as the \emph{ground truth}. From this ``true'' prior, we perform forward simulations: sampling first from the prior, then from the likelihood, and carrying out all necessary computational steps to generate a set of model-implied statistics corresponding to the ground truth.
This set of simulated statistics can then serve as a proxy for the ``expert-elicited'' statistics. 

This \emph{oracle} approach allows us to assess the self-consistency of the algorithm, the degree of non-identifiability (i.e., how informative are the elicited statistics), and the suitability of the chosen algorithm parameters used for training NFs with mini-batch SGD (such as the learning rate or the number of epochs). 
However, there is one important caveat when using an \emph{oracle}: the user might be tempted to compare the learned prior distributions against the true prior distributions (which are now available). Specifically, if there is an unsatisfactory match between the true and learned prior distributions, the user might interpret this mismatch as a failure in learning. It is important to remember, however, that the method only knows about the elicited statistics, which are incorporated into the loss function. Thus, when the elicited statistics do not incorporate information about the model parameters but instead pertain to observable quantities, there is, from the method's perspective, no concept of a true prior. In particular, if the true and learned elicited statistics match but there is a mismatch in the learned prior, this is likely a result of an underidentified problem, and the objective function requires additional information to properly constrain the problem.

\section{Implementation}\label{subsec:sbt}
In all simulation studies, we use a batch size of $128$. 
To simulate from an oracle (discussed in Section~\ref{subsec:general-setup}), we draw once $S=10,000$ samples from the true joint prior and then compute the corresponding set of ``expert''-elicited statistics.
During training with mini-batch SGD we draw $S=200$ samples from the joint prior.
To evaluate the discrepancy between the model-implied and expert-elicited statistics, we use a squared, biased \emph{Maximum Mean Discrepancy} (MMD) loss with an energy kernel \citep{gretton2006kernel, feydy2019interpolating}, following our previous success with this metric \citep{bockting2023simulation}.
For the correlation between parameters, however, we apply a squared-error loss (L2 loss; for details see Section~\ref{subsec:general-setup}). The total loss is computed as a weighted sum of the individual loss components. The values of the respective weights are discussed together with the selection of the elicited statistics whereby each elicited statistic represents one loss component (see Section~\ref{subsec: selection-of-target-quantities}).

The architecture of the NFs is rather simple compared to typical applications in deep learning \citep[e.g.,][]{kingma2018glow} and consists of a standard multivariate Gaussian as a base distribution and three affine coupling blocks. Each coupling block consists of two dense layers with $128$ units and ReLU activation functions. In Simulation Study 1, the number of learnable hyperparameters is $|\lambda| = 202,776$, while in the case studies corresponding to Simulation Study 2, it is $|\lambda| = 205,872$.

In Simulation Study 1, we assume a discrete likelihood and use the softmax-gumble trick to compute the gradients of a discrete random variable \citep[for more information, see][]{bockting2023simulation}. 
The gradients of the total loss with respect to the hyperparameters are used to update the hyperparameters $\lambda$. Optimization was performed with the Adam optimizer for $500$ epochs in Simulation Study 1 and $800$ epochs in Simulation Study~2. We used a fixed learning rate of $0.00025$ for Simulation Study 2, Scenario 1 \& 2, and $0.0001$ for Simulation Study 1 and Stimulation Study 2, Scenario 3.
These algorithm (hyper-)parameter values were obtained by manual tuning until good performance across all of our case studies was achieved.

All simulations and plots were performed in Python 3.11 using our Python package \emph{elicito} 0.3.1 \citep{Bockting_elicito}, which relies, among others, on the following libraries: TensorFlow 2.14 \citep{abadi2016tensorflow}, TensorFlow Probability 0.23 \citep{dillon2017tensorflow}, NumPy 1.26 \citep{harris2020array}, pandas 2.2.3 \citep{reback2020pandas}, and BayesFlow 1.1.6 \citep{radev2023bayesflow}. To ensure the reproducibility of our results, the specific version of \emph{elicito} used in this paper has been archived on Zenodo and is accessible via the following DOI \url{https://doi.org/10.5281/zenodo.15241853}. Simulations were executed on a YOGA Pro 9i laptop equipped with GPU acceleration.

\section{Simulation Studies}\label{sec:simulation-studies}
\subsection{General Setup}\label{subsec:general-setup}
In the following, we present four case studies organized across two simulation studies to evaluate the performance of our elicitation method. The selection of case studies is motivated by the desire to choose scenarios that are simple and allow us to study the behavior of the method in a well-controlled setting. Since this paper presents the extension with NFs for the first time, it is important for us to first develop an understanding of the method’s behavior. In future work, we will examine the behavior of our method in more complex models that are closer to real-world applications.

In Simulation Study 1, we introduce a simple Binomial regression model with one continuous predictor. In Simulation Study 2, we present three scenarios based on a normal regression model with a three-level categorical predictor. The three normal scenarios differ only in the pre-defined joint prior representing the ground truth. Table~\ref{tab:overview-models} provides an overview of the generative models and the joint priors representing the ground truth used in the simulation studies. Before introducing the simulation studies, we will discuss the decisions made in the simulation setup in greater detail.
All code and results can be found on GitHub (\url{https://github.com/florence-bockting/non-parametric-prior}) and OSF (\url{https://osf.io/xrzh6/}), respectively. Information about the training time for each simulation study is provided in Appendix~\ref{subsecA1}.

\begin{table}[ht]
    \centering
    \begin{tabular}{p{0.24\textwidth} p{0.24\textwidth} p{0.43\textwidth}}
    \hline 
    \\[-0.8em]
    \emph{Model}  & \emph{Generative Model} & \emph{Ground Truth} \\[0.4em]
    \hline
    \\[-0.8em]
    Binomial --- M1 & $y_i \sim \text{Binomial}(p_i, 30)$ & $\beta_0 \sim \text{Normal}(0.1, 0.1)$ \\
    (Section~\ref{subsec:binomial}) & $p_i = \text{sigmoid}(\beta_0+\beta_1x_i)$ & $\beta_1 \sim \text{Normal}(-0.1, 0.3)$ \\[0.4em]
    \hline
    \\[-0.8em]
    \textbf{Normal models} &  & \\
    \quad \emph{Scenario 1} --- M2 & $y_i \sim \text{Normal}(\mu_i, \sigma)$ & $\beta_0 \sim~\text{Normal}(10, 2.5)$ \\
     \quad (Section~\ref{subsec:independent-normal}) & $\mu_i = \beta_0+\beta_1 x_{1,i}+\beta_2 x_{2,i}$ & $\beta_1 \sim \text{Normal}(7, 1.3)$\\
     && $\beta_2 \sim \text{Normal}(2.5, 0.8)$ \\
     && $\sigma \sim \text{Gamma}(5, 2)$ \\[0.4em]
     \hdashline
     \\[-0.8em]
     \quad \emph{Scenario 2} --- M3 & see M2 & $\beta_0 \sim~\text{Normal}(10, 2.5)$\\
     \quad (Section~\ref{subsec:skewed-normal}) && $\beta_1 \sim \text{SkewNormal}(7, 1.3, 4)$\\
     && $\beta_2 \sim \text{SkewNormal}(2.5, 0.8, 4)$ \\
     && $\sigma \sim \text{Gamma}(5, 2)$ \\[0.4em]
     \hdashline 
     \\[-0.8em]
     \quad \emph{Scenario 3} --- M4 \\[-1.3em] \quad (Section~\ref{subsec:correlated-normal})& see M2 & $\beta \sim \text{Mv-Normal}\left(\begin{bmatrix}  10 \\ 7 \\ 2.5\end{bmatrix}, \; \text{D}(s) \, \textbf{R} \, \text{D}(s)\right)$ \\
    \\
      && $\textbf{R} = \begin{bmatrix} 1. & 0.3 & -0.3 \\
            0.3 & 1. & -0.2 \\ -0.3 & -0.2 & 1. \end{bmatrix}$\\
    \\
    && $s = ( 2.5 , 1.3 , 0.8 )$\\
    && $\sigma \sim \text{Gamma}(5, 2)$ \\
    \hline
    \end{tabular}
    \caption{Overview of generative models and ground truth used in each simulation study. Symbols: $\mathbf{R}$ denotes the correlation matrix and $s$ the vector of standard deviations.}
    \label{tab:overview-models}
\end{table}

\paragraph{Simulated Expert as Ground Truth}
For the simulation studies in this paper, we simulate expert input using synthetic data generated from a predefined joint prior representing the ground truth (i.e., oracle, see Section~\ref{subsec: evaluation-results} for details). For simplicity, we use a parametric prior with fixed hyperparameter values as ground truth for each case study. 
Further details on the exact specification of the ``true'' joint prior is provided in the respective section and in Table~\ref{tab:overview-models}.

\paragraph{Selection of Target Quantities and Elicitation Techniques}
\label{subsec: selection-of-target-quantities}
Following recommendations in the field, we tried to select mainly observable quantities as target quantities for our simulation studies (see Section~\ref{subsec: target-quantities-elicitation-techniques} for details). 
For the Binomial regression model, we used the expected observations of the outcome variable conditional on two observations of the continuous predictor, $y \mid x_0$ and $y \mid x_1$. For the normal regression model, we considered the prior predictive observations conditional on three groups of the categorical predictor, $y \mid \text{gr}_i$ for $i=1,2,3$.

However, we had to deviate from the recommendation to select observable target quantities in two respects.
First, we included $R^2$ as an additional target quantity for the normal model (Simulation Study 2), motivated by the following reasoning: if we only have information about the outcome variable for each of the three groups, we provide the learning algorithm with no means to differentiate between parameter uncertainty and data uncertainty. The coefficient of determination, $R^2$, represents the ratio of parameter uncertainty to total uncertainty (which encompasses both data and parameter uncertainty). Consequently, it offers more information for the learning algorithm to differentiate between these two types of uncertainty, even though it may be somewhat more challenging for a domain expert to interpret. However, as $R^2$ is commonly reported as a metric in a regression context, we would argue that this quantity is still well understood by domain experts.

Second, we included the pairwise correlation between model parameters in the set of target quantities for each case study. If independence between model parameters is assumed, all correlations are set to zero; otherwise, we use the exact correlation structure indicated by the ground truth. We acknowledge that this information cannot be reasonably demanded from a domain expert in most cases. However, we currently lack suitable elicitation methods for asking domain experts about interpretable quantities that provide sufficient information regarding the correlation structure between the model parameters.
A future task will be to develop such elicitation methods and to then incorporate these methods into our computational framework.

Regarding the selection of \emph{elicitation techniques}, we follow the recommendation to query quantiles from a domain expert \citep{kadane1998experiences}. For each target quantity, except for the correlation information, we elicit five quantiles: $5\%,25\%,50\%,75\%,95\%$. For the correlation values, we assume that the expert provides a single point estimate corresponding to a moment-based elicitation approach.

In summary, the elicited statistics for Simulation Study 1 consist of five quantiles each for $y \mid x_0$ ($t_1$) and $y \mid x_1$ ($t_2$), as well as the correlation information $\rho(\beta_0, \beta_1)$ ($t_3$). Each of these three elicited statistics corresponds to a distinct loss component in the objective function. The elicited statistics for Simulation Study 2 consist of five quantiles each for $R^2$ ($t_1$) and $y\mid \text{gr}_i$ with $i=1,2,3$ ($t_2, t_3, t_4$), as well as the correlation information between the model parameters ($t_5$).
The discrepancy in the correlation values is measured using an L2 loss, whereas the discrepancies for the remaining components are computed using the $\text{MMD}^2$ loss. To account for differences in scale—arising both from the use of different discrepancy measures and from the varying domains of the elicited statistics—we weight the individual loss components as follows: 
the correlation loss component is scaled by a factor of $0.1$, the $R^2$ loss component by a factor of $10.0$, and all other loss components are uniformly weighted with a factor of $1.0$.

To assess whether the selected set of elicited statistics is sufficiently informative to determine a prior distribution for the model parameters, we conduct a \emph{sensitivity analysis} for each case study.

\paragraph{Evaluation of Simulation Results}\label{subsec:evaluation-results}
We run each simulation study 30 times with different random seeds. 
As preliminary convergence checks, we calculate the slope of the loss trajectory over the last 100 epochs and perform a visual convergence check for the five replications with the highest absolute slopes (i.e., worst cases).
Additionally, we examine the trajectories of the total loss, the individual loss components related to each elicited statistic, as well as the mean and standard deviation for each marginal prior across epochs. 
Furthermore, to visually assess whether the learning criterion has been accurately captured, we compare the final model-implied statistics with the expert-elicited statistics. 
Finally, we examine the learned prior distributions for each replication and compute the model average across the 30 learned prior distributions. Table~\ref{tab:analysis_workflow} provides a summary of the analysis workflow followed in each of the subsequent simulation studies.
\begin{table}[h!]   
    \begin{tabular}{p{0.24\textwidth} p{0.24\textwidth} p{0.43\textwidth}}
        \hline \\[-0.5em]
         \emph{Task} & \emph{Approach} & \emph{Insights} \\[0.3em]
         \hline
         \multicolumn{3}{l}{\textbf{Setup \& Elicitation Stage}}\\
         \hline
         Select the set of elicited statistics and evaluate its informativeness. & (1) Simulate elicited statistics by defining a true joint prior (i.e., the oracle). (2) Conduct a sensitivity analysis. & Can the set of elicited statistics provide sufficient information to learn the key aspects of the joint prior? \\
         \hline
         \multicolumn{3}{l}{\textbf{Fitting Stage}}\\
         \hline
         Evaluate variability in learned priors to assess non-identifiability. & Run the training algorithm several times with different random seeds. & Do different training runs yield different results?\\
         \hdashline
         Verify convergence of the training algorithm. & (1) Examine the slope of final loss values across replications. (2) Visually check the convergence of key quantities for selected seeds. & Will training the model for additional epochs lead to a practically relevant improvement in performance or a reduction in total loss? \\
         \hdashline
         Verify accurate learning of expert data. & Visually compare model-implied and expert-elicited statistics. & Was the method successful in accurately capturing the expert data? \\
         \hline
         \multicolumn{3}{l}{\textbf{Evaluation Stage}}\\
         \hline
         Examine the learned priors across the different seeds & Visually inspect the marginal priors and correlations between model parameters. & How much do the learned priors vary between different seeds? Were any degenerate prior distributions learned? Do we observe any form of multi-modality? Can any of the learned priors be ruled out as unreasonable post hoc? \\
         \hdashline
         Obtain the final learned prior for subsequent analysis. & Perform model averaging using weights based on relative loss performance. & Instead of selecting a single prior, compute a model average over practically equivalent priors to account for additional variability due to non-identifiability. \\
         \hline
    \end{tabular}
    \caption{Overview of the analysis workflow used in the simulation studies.}
    \label{tab:analysis_workflow}
\end{table}

\paragraph{Additional simulation results}
For each case study, we conduct two additional sets of simulations, which are not part of the main analysis but are included in the supplementary material for completeness and further comparison. The supplementary material can be found on GitHub at \url{https://github.com/florence-bockting/non-parametric-prior}. As these simulations are complementary to the main focus of the paper, we do not report the results in the main text.

In the first additional set of simulations, we re-ran all case studies using the \emph{model parameters} as the ``elicited statistics'' (instead of an indirect quantity derived from the model parameters). In this case, the method trains directly on the model parameters, enabling us to assess whether it is, in principle, capable of recovering the true joint prior distribution when complete information is available. Thus, this set of simulations serves as a validation set for our method.

In the second additional set of simulations, we applied the parametric-prior method proposed in \citet{bockting2023simulation} to all case studies. This allows for a direct comparison between the performance of the proposed non-parametric approach and the previously introduced parametric variant within the same simulation-based framework. The only exception is Scenario 3 of Simulation Study 2, where we introduce a multivariate normal prior. In this case, a comparison is currently not possible, as the optimization of a full covariance matrix is not yet implemented in \emph{elicito} \citep{Bockting_elicito}, but it is planned for a (near) future release.

\subsection{Simulation Study 1: Binomial Likelihood with Independent Normal Priors}\label{subsec:binomial}
We introduce the general idea of our method using a Binomial model with a logit link (sigmoid response function) and a single continuous predictor, $x$. The predictor was generated by first constructing a sequence from 1 to 50, which was then scaled by its standard deviation. In the following we refer to this model as \ref{M1}:
\begin{align}\label{M1}
    \tag{M1}
    \begin{split}
        y_i &\sim \text{Binomial}(p_i, 30)\\
        p_i &= \text{sigmoid}(\beta_0+\beta_1x_i)\\
        \beta_0, \beta_1 &\sim p_\lambda(\beta_0, \beta_1) \\
        \theta &\equiv (\beta_0,\beta_1) 
    \end{split}
\end{align}
The goal is to learn a joint prior for the model parameters $\beta_0$ (intercept) and $\beta_1$ (slope), assuming independence between these parameters. 
\paragraph{Setup \& Elicitation Stage}
As elicited statistics, we select prior predictions for two values of the continuous predictor, $y \mid x_0$ and $y \mid x_1$. The two values of $x$ correspond to the $25\%$ and $75\%$ quantiles of the predictor. We select specific quantiles rather than randomly sampling two observations from the predictor to avoid the observations being too close together, which would reduce informativeness. 
From the distribution of the two selected observations, we compute five quantiles which represent the elicited statistics (see Section~\ref{subsec:general-setup} for details). Note that $y \mid x_0$ and $y \mid x_1$ each have an associated distribution, resulting from simulating $y^{(s)}$ for $S$ samples drawn from the joint prior and conditioning on the respective $x$ value. The distribution over $y \mid x$ encodes the parameter uncertainty.

To obtain the ``expert''-elicited statistics, we define a true prior that represents the ground truth and simulate from the generative model in forward mode, computing the corresponding true-elicited statistics. The \emph{true} joint prior is defined by independent normal distributions for each model parameter: $\beta_0 \sim \text{Normal}(0.1, 0.1)$ and $\beta_1 \sim \text{Normal}(-0.1, 0.3)$. 

\begin{figure}[ht]
    \centering
    \includegraphics[width=0.6\linewidth]{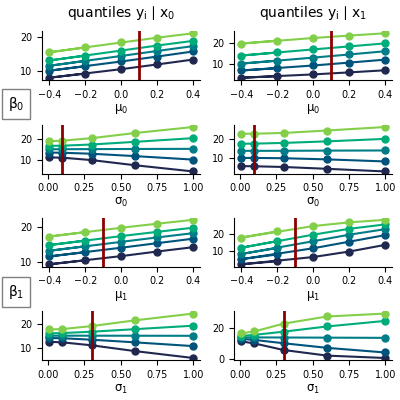}
    \caption{\emph{Binomial model---Sensitivity Analysis}: Rows represent hyperparameters of each model parameter. Columns represent elicited statistics: five quantiles for $y \mid x_0$ and $y \mid x_1$. Quantiles are depicted in different colors. In each row, the corresponding hyperparameter, is varied across the range shown on the x-axis, while all other hyperparameters are held constant at their true values, indicated by the red vertical line. The upper two rows indicate results for the intercept parameter $\beta_0$ and the lower two rows for the slope parameter $\beta_1$.}
    \label{fig:sensitivity_m1}
\end{figure}

To assess the \emph{informativeness} of the selected set of elicited statistics, we conduct a sensitivity analysis, with the results shown in Figure~\ref{fig:sensitivity_m1}. In each row, a single hyperparameter is varied across the range shown on the x-axis, while all other hyperparameters are held constant at their true values, indicated by the red vertical line. The columns present the two elicited statistics, specifically the five quantiles (in shades of blue and green) for $y \mid x_0$ and $y \mid x_1$. The upper two rows show results for the intercept parameter $\beta_0$ and the lower two rows for the slope parameter $\beta_1$.
The results of the sensitivity analysis indicate that the selected elicited statistics provide sufficient informativeness for the model hyperparameters. 

\begin{figure}
    \centering
    \includegraphics[width=.7\linewidth]{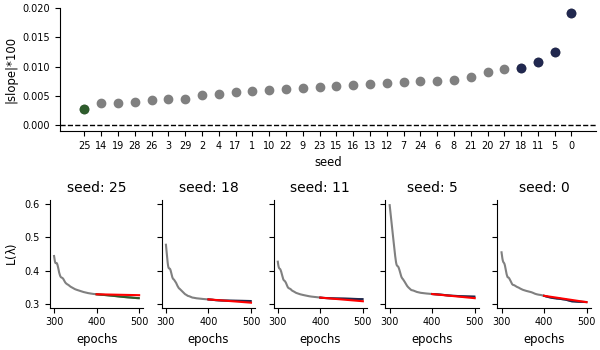}
    \caption{\emph{Binomial model---Preliminary convergence check for all 30 replications}: Upper plot: Absolute slope of the total loss trajectory (scaled by a factor of 100) over the last 100 epochs on the y-axis, with each replication shown along the x-axis. Replications (i.e., seeds) are arranged in ascending order based on the magnitude of the absolute slope. Best-performing model is seed 25, while the worst-performing models are seeds 18, 11, 5, and 0. Lower plot: Loss trajectories of the best- and worst-performing seeds over the last 200 epochs.}
    \label{fig:convergence-check-binom-loss}
\end{figure}

\paragraph{Fitting Stage}
We ran the training algorithm 30 times with different random seeds and conducted an initial convergence check by inspecting the slope of the last 100 epochs for each replication. In Figure~\ref{fig:convergence-check-binom-loss}, the upper row shows the absolute slope (scaled by a factor of 100) for each replication. In the lower row, the learning trajectory of the total loss is shown for the best-performing (leftmost) and the four worst-performing replications. We observe a near-zero slope for the best model, indicating convergence, while the negative slopes for the worst-performing seeds suggest that the learning algorithm continues to make progress.

\begin{figure}[!ht]
    \centering
    \includegraphics[width=0.6\linewidth]{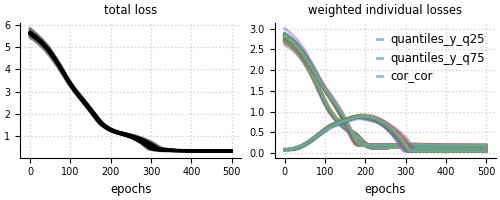}
    \caption{\emph{Binomial model---Visual convergence check 1}: Updating trajectory across epochs for all 30 replications. On the left, trajectory of the total loss and on the right, trajectory of the weighted individual loss components.}
    \label{fig:visual-loss-convergence-binom}
    \includegraphics[width=0.65\linewidth]{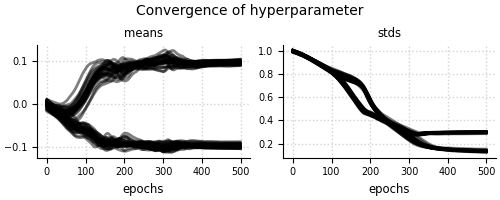}
    \caption{\emph{Binomial model---Visual convergence check 2}: Updating trajectory across epochs for for all 30 replications. Trajectory of the mean (left plot) and standard deviation (right plot) of the marginal priors.}
    \label{fig:convergence-check-binom-hyperparameter}
\end{figure}

However, a visual inspection of the learning trajectories of the total loss and individual loss components (Figure~\ref{fig:visual-loss-convergence-binom}) and the means and standard deviations of the marginal priors (Figure~\ref{fig:convergence-check-binom-hyperparameter}), indicates satisfactory convergence. Therefore, although the slope analysis indicates minor ongoing learning progress, no substantial changes in the results are expected if the model is trained for additional epochs. Overall, we conclude that the training process was successful.

\begin{figure}[ht]
    \centering
    \includegraphics[width=.65\linewidth]{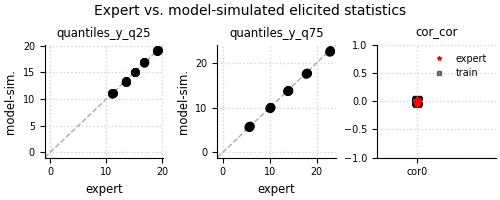}
    \caption{\emph{Binomial model---Learned elicited statistics for all 30 replications}: The two leftmost plots show the learned quantiles (y-axis) vs. the true quantiles (x-axis) of each replication for the two target quantities, $y \mid x_0$ and $y \mid x_1$. The diagonal line serves as a reference, where points lying on the line indicate a perfect match between learned and true quantiles. The rightmost plot displays the learned (black) vs. true (red) correlations. Given that we assume independence between model parameters in this scenario, the true correlation is zero.}
    \label{fig:elicited-statistics-binom}
\end{figure}

The corresponding final learned elicited statistics for each replication are shown in Figure~\ref{fig:elicited-statistics-binom}. In the first two plots, each true quantile (x-axis) is plotted against its corresponding learned quantile (y-axis) for each replication. Points that align perfectly with the diagonal dashed line represent a perfect match between true and learned quantiles. Points above the diagonal indicate that the learned quantiles are higher than the true ones (and vice versa for lower points). Accuracy of the learned correlation information is presented in Figure~\ref{fig:elicited-statistics-binom}. The true (red) correlation matches perfectly with the learned (black) correlations.

\begin{figure}[ht]
    \centering
    \includegraphics[width=.8\linewidth]{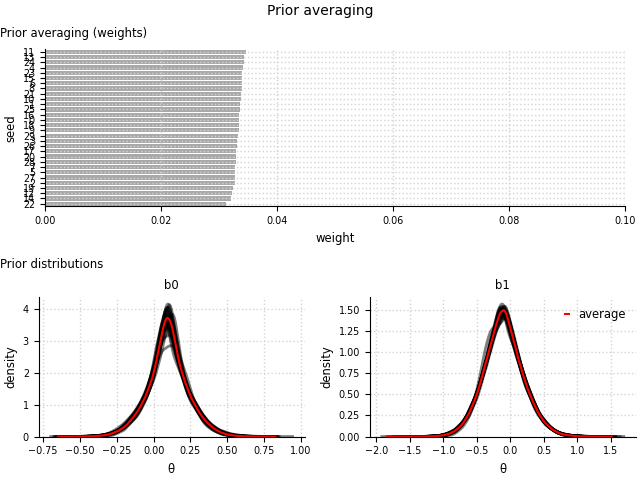}
    \caption{\emph{Binomial model---Learned marginal priors for all 30 replications}: Upper plot: Computed weights per replication used for model averaging, reflecting the relative performance of each model based on the final total loss. In this scenario, all replications perform equally well, resulting in almost uniform weights. Lower plot: The learned marginal priors for each replication and in red the prior average. In this case, the elicited statistics are sufficiently informative to effectively identify the generative model, resulting in only minor variations in the learned priors across replications.}
    \label{fig:priors-binom}
\end{figure}

\paragraph{Evaluation Stage}
Finally, we examine the learned prior distributions. Figure~\ref{fig:priors-binom} shows the learned marginal priors for each replication, alongside the results from the model averaging approach. The weights used for model averaging are shown in the upper plot of Figure~\ref{fig:priors-binom} and reflect the relative performance of each replication based on their total loss (see Section~\ref{subsec:evaluation-results} for details). In this case, all replications perform similarly well, resulting in an almost uniform weighting scheme. The resulting averaged prior is represented by the solid red line in the lower plot, shown together with the learned priors from each replication.

\subsection{Simulation Studies 2: Normal Likelihood}\label{sec:normal-scenarios}
In the remaining simulation studies, we aim to evaluate the method's performance using a model with slightly increased complexity, specifically a normal regression model with standard deviation $\sigma$ and mean $\mu_i$, where $\mu_i$ depends on a three-level grouping variable encoded via the dummy variables $x_{1}$ and $x_{2}$:
\begin{align*}
    \begin{split}
        y_i &\sim \text{Normal}(\mu_i, \sigma)\\
        \mu_i &= \beta_0+\beta_1 x_{1,i}+\beta_2 x_{2,i}\\
        \beta_0, \beta_1,\beta_2, \sigma &\sim p_\lambda(\beta_0, \beta_1, \beta_2, \sigma) \\
        \theta &\equiv (\beta_0,\beta_1,\beta_2,\sigma) 
    \end{split}
\end{align*}
with $\beta_0$, $\beta_1$, and $\beta_2$ being the regression coefficients representing the intercept and the two group contrasts, respectively.
The goal is to learn a joint prior for the model parameters $\theta$.
To systematically assess our method, we use the same model across all upcoming simulation studies, varying only the \emph{true} joint prior in each scenario:
\begin{itemize}
    \item \textbf{Scenario 1 (M2)}: Independent normal priors for the regression coefficients are defined as $\beta_0 \sim~\text{Normal}(10, 2.5), \beta_1 \sim \text{Normal}(7, 1.3), \beta_2 \sim \text{Normal}(2.5, 0.8)$, with a Gamma prior for the random noise, $\sigma \sim \text{Gamma}(5, 2)$ 
    \item \textbf{Scenario 2 (M3)}: Identical to M2 but introduces skewed normal priors for $\beta_1$ and $\beta_2$, where $\beta_1 \sim \text{SkewNormal}(7, 1.3, 4)$\footnote{For the skewed normal distribution, we use the implementation from TensorFlow Probability \citep{abadi2016tensorflow}, which features the two-piece normal distribution \citep{fernandez1998bayesian}. This distribution is parameterized by location, scale, and shape parameters. The shape parameter controls the skewness, with values above one resulting in a right-skewed distribution.} and $\beta_2 \sim \text{SkewNormal}(2.5, 0.8, 4)$.
    \item \textbf{Scenario 3 (M4)}: Identical to M2 but introduces a dependency structure among the model coefficients $\beta$ using a correlated multivariate normal distribution, with marginal distributions identical to those in M2. The correlated prior is specified as:
    \begin{align*}
            \beta&\sim \text{Mv-Normal}\left(\begin{bmatrix}  10 \\ 7 \\ 2.5\end{bmatrix}, \; \text{D}(s) \, \textbf{R} \, \text{D}(s)\right) \quad \text{with}\\
            \textbf{R} &= \begin{bmatrix} 1. & 0.3 & -0.3 \\
            0.3 & 1. & -0.2 \\ -0.3 & -0.2 & 1. \end{bmatrix}, \quad
            s = ( 2.5 , 1.3 , 0.8 ) 
    \end{align*}
    where $\mathbf{R}$ denotes the correlation matrix and $s$ the vector of standard deviations.
\end{itemize}

\subsubsection{Scenario 1: Independent Normal Priors}
\label{subsec:independent-normal}
\begin{figure}[ht]
    \centering
    \includegraphics[width=1.\linewidth] {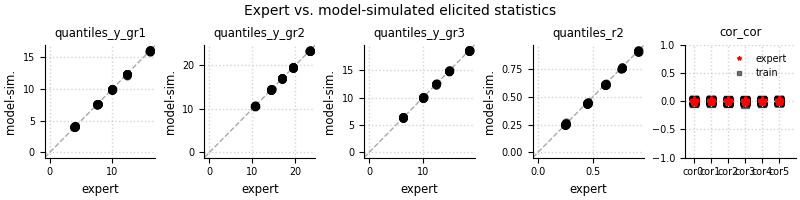}
    \caption{\emph{Normal model / Scenario 1 (M2) --- Learned elicited statistics for all 30 replications}: First three plots (from the left) show the learned quantiles (y-axis) vs. the true quantiles (x-axis) of each replication for the four target quantities $y \mid \text{gr}_i$ with $i=1,2,3$ and $R^2$. The diagonal line serves as a reference, where points lying on the line indicate a perfect match between learned and true quantiles. The rightmost plot displays the learned (black) vs. true (red) correlations between the model parameters. Given the independence assumption, the true correlation is zero.}
    \label{fig:elicited-statistics-independent-normal}
\end{figure}

\begin{figure}[t]
    \centering
    \includegraphics[width=.8\linewidth] {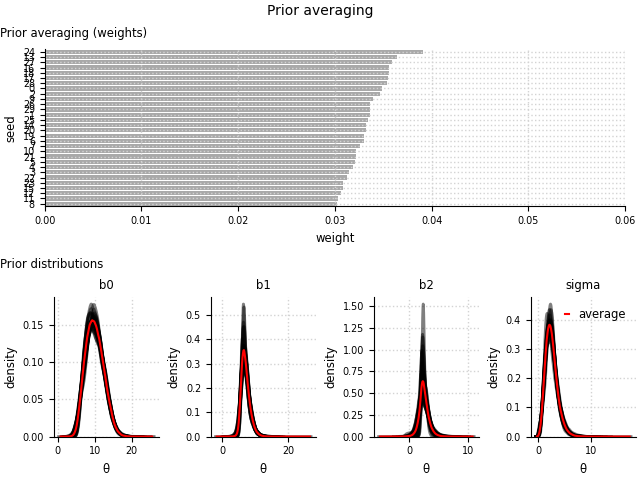}
    \caption{\emph{Normal model / Scenario 1 (M2) --- Learned marginal priors across replications}: Upper plot: Computed weights per replication used for model averaging, reflecting the relative performance of each model based on the final total loss. Lower plot: The learned marginal priors for each replication (in black) and the prior average (in red). }
    \label{fig:marginals-independent-normal}
\end{figure}

\paragraph{Setup \& Elicitation Stage}
We begin with a brief assessment of the selected elicited statistics, with sensitivity analysis results detailed in Appendix~\ref{subsecA2}. 
While the intercept coefficient, $\beta_0$, is well-informed by all three target quantities related to the predictive distributions of the groups, the slope coefficients, $\beta_1$ and $\beta_2$, are each informed by only one target quantity, specifically those corresponding to group 2 and group 3, respectively. Consequently, we anticipate greater difficulty in accurately learning these coefficients.
Additionally, we included $R^2$ among the target quantities to aid the learning algorithm in distinguishing parameter uncertainty from data uncertainty. 
\paragraph{Fitting Stage}
To verify successful convergence, we first examined the absolute slopes across all 30 replications, followed by a detailed visual inspection of the learning trajectories for the loss components and additional quantities of interest (see Appendix~\ref{subsecA2} for results). The analysis indicates that the learning process was successful. The model-implied statistics are shown in Figure~\ref{fig:elicited-statistics-independent-normal} and demonstrate a strong match between the learned and true elicited statistics, supporting the assumption of successful convergence.
\paragraph{Evaluation Stage}
The corresponding learned marginal priors of each replication alongside with the model averaging results are depicted in Figure~\ref{fig:marginals-independent-normal}. 
In this scenario, we observe noticeably greater variation in the learned priors across different replications, particularly for $\beta_1$ and $\beta_2$. This variation is also reflected in the model averaging weights, which now deviate clearly from the uniform distribution observed in Simulation Study 1.
These results suggest that while the elicited statistics provide sufficient information to learn the marginal priors of $\beta_0$ and $\sigma$, they are insufficient to uniquely identify the priors for $\beta_1$ and $\beta_2$.
It is important to emphasize that the only information available to the method for learning the prior distributions is based on the elicited statistics shown in Figure~\ref{fig:elicited-statistics-independent-normal}. Given this limited set of information, learning a non-parametric joint prior becomes a highly underconstrained problem, where multiple prior distributions can be consistent with the same set of elicited statistics.
However, this should not be interpreted as a limitation of the method itself, but rather as a consequence of the limited information available to the method for learning the joint prior.

\subsubsection{Scenario 2: Skewed Normal Priors}\label{subsec:skewed-normal}
\begin{figure}[ht]
    \centering
    \includegraphics[width=1.\linewidth] {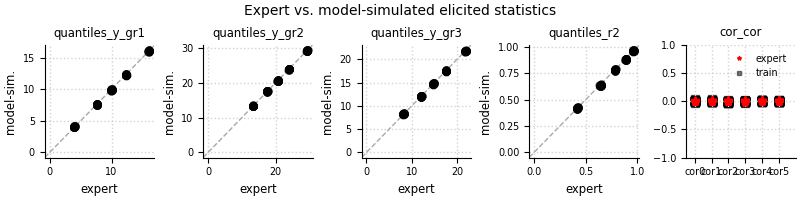}
    \caption{\emph{Normal model / Scenario 2 (M3) --- Learned elicited statistics for all 30 replications}: First four plots from the left depict the learned quantiles (y-axis) vs. the true quantiles (x-axis) of each replication for the four target quantities $y \mid \text{gr}_i$ with $i=1,2,3$ and $R^2$. The diagonal line serves as a reference, where points lying on the line indicate a perfect match between learned and true quantiles. The rightmost plot displays the learned (black) vs. true (red) correlations between the model parameters. Given the independence assumption, the true correlation is zero.}
    \label{fig:elicited-statistics-skewed-normal}
\end{figure}

\begin{figure}[ht]
    \centering
    \includegraphics[width=0.8\linewidth] {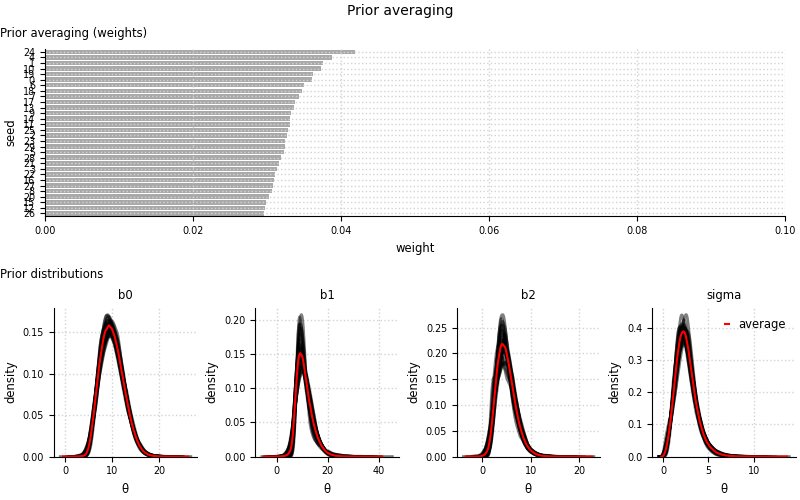}
    \caption{\emph{Normal model / Scenario 2 (M3) --- Learned marginal priors for all 30 replications}: Upper plot: Computed weights per replication used for model averaging, reflecting the relative performance of each model based on the final total loss. Lower plot: The learned marginal priors for each replication (in black) and the prior average (in red).}
    \label{fig:marginals-skewed-normal}
\end{figure}

\paragraph{Setup \& Elicitation Stage}
Scenario 2 differs from the previous one only in the ground truth, where we introduce a skew-normal marginal prior for the two slope coefficients, $\beta_1$ and $\beta_2$. In this scenario, we examine whether the method can learn this additional skewness property using the same ``elicited statistics'' as those introduced in the previous case study. 
The sensitivity analysis (see Appendix~\ref{subsecA3}) already suggests that there is likely insufficient information to disentangle the variance and skewness components for the two slope coefficients; therefore, we except to observe more variation in the learned prior distributions.
\paragraph{Fitting \& Evaluation Stage}
The convergence diagnostics are depicted in Appendix~\ref{subsecA3} and indicate successful convergence. The corresponding match between the model-implied and true ``elicited'' statistics is shown in Figure~\ref{fig:elicited-statistics-skewed-normal}, demonstrating overall good performance.
Figure~\ref{fig:marginals-skewed-normal} shows the corresponding learned marginal priors together with the results from prior averaging. The results indicate relatively stable learning of the prior distributions across all replications. As before, the priors for the slope coefficients exhibit higher variation. Most importantly, however, the method is able to learn the introduced skewness. 

\subsubsection{Scenario 3: Correlated Normal Priors}\label{subsec:correlated-normal}
\begin{figure}[ht]
    \centering
    \includegraphics[width=1.\linewidth] {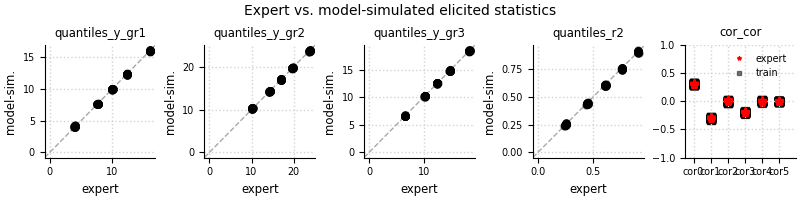}
    \caption{\emph{Normal model / Scenario 3 (M4) --- Learned elicited statistics for all 30 replications}: First four plots show the learned quantiles (y-axis) vs. the true quantiles (x-axis) of each replication for the four target quantities $y \mid \text{gr}_i$ with $i=1,2,3$ and $R^2$. The diagonal line serves as a reference, where points lying on the line indicate a perfect match between learned and true quantiles. The rightmost plot displays the learned (black) vs. true (red) correlations between the model parameters.}
    \label{fig:elicited-statistics-correlated-normal}
\end{figure}

In this final scenario, we introduced a dependency structure among the regression coefficients, while all other aspects remained identical to Scenario 1 (\ref{M1}). The results of the sensitivity analysis and the convergence checks are provided in Appendix~\ref{subsecA4} and show successful convergence.
The model-implied and true ``elicited'' statistics are shown in Figure~\ref{fig:elicited-statistics-correlated-normal} and indicate a close match for all elicited statistics.

\begin{figure}[ht]
    \centering
    \includegraphics[width=.8\linewidth] {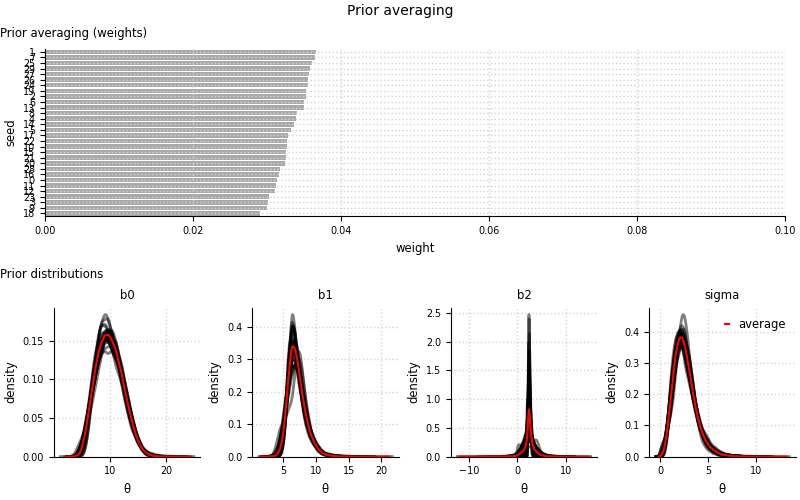}
    \caption{\emph{Normal model / Scenario 3 (M4) --- Learned marginal priors for all 30 replications}: Upper plot: Computed weights per replication used for model averaging, reflecting the relative performance of each model based on the final total loss. Lower plot: The learned marginal priors for each replication (black) and the prior average (red).}
    \label{fig:marginals-correlated-normal}
\end{figure}

\begin{figure}[ht]
    \centering
    \includegraphics[width=0.6\linewidth]{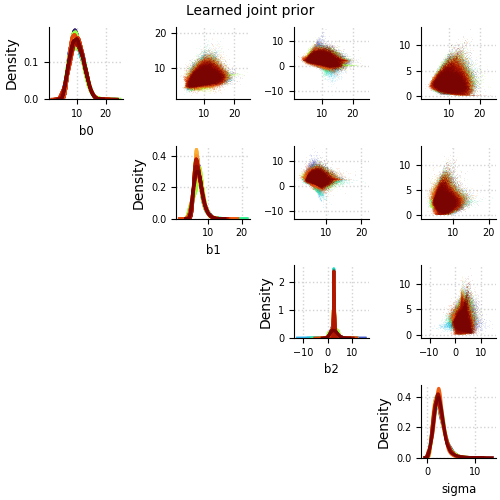}
    \caption{\emph{Normal model / Scenario 3 (M4) --- Learned joint prior for all 30 replications} Each replication is shown in a different color. The correlation structure of the regression coefficients used for learning the prior is $\rho(\beta_0,\beta_1)=0.3, \rho(\beta_1,\beta_2)=-0.2,$ and $\rho(\beta_0,\beta_2)=-0.3$. Independence between $\sigma$ and correlation coefficients is assumed.}
    \label{fig:joint-correlated-normal}
\end{figure}

In Figure~\ref{fig:marginals-correlated-normal}, the final learned marginal priors for each replication are shown alongside the results from prior averaging. Results show substantial variation in the learned priors for $\beta_2$, whereas the priors for all other parameters show relatively high consistency across replications. Since the focus of this case study is specifically on the correlation structure of the regression coefficients, it is informative to also examine the joint prior distribution, which is shown in Figure~\ref{fig:joint-correlated-normal}. The joint priors from different replications are shown in different colors. Although there is clear variation in the learned joint priors, the method is able to capture the underlying correlational pattern. Although this result is not particularly surprising, as we provided the method with the exact correlation information between the model parameters, it nonetheless indicates that the method is capable of learning the correlational structure of a joint prior when provided with the relevant information. Therefore, future work is needed to develop elicitation methods that enable the method to obtain relevant information by querying domain experts indirectly.

\section{Discussion \& Outlook}\label{sec:discussion}
In this paper, we introduce an extension of our simulation-based framework \citep{bockting2023simulation} for learning \emph{non-parametric joint priors}. This extension builds on using normalizing flows \citep{kobyzev2020normalizing} to learn flexible joint priors for the model parameters. Although we employ normalizing flows in this work, other generative models, such as flow matching \citep{lipman2022flow} or diffusion models \citep{cao2024survey} are also compatible with our framework and can be readily used as a drop-in replacement. 

Across four simulation studies, we demonstrated successful learning of joint priors based primarily on interpretable quantities that can be meaningfully provided by domain experts. One exception is the use of correlation information, which the current method requires to reasonably constrain the optimization problem. Future work is required to integrate in our method an elicitation technique that infers the correlation indirectly from the expert by asking about interpretable quantities. Recently, \citet{mikkola2024preferential} proposed a method that uses pairwise comparisons or ranking techniques to learn a multivariate prior density, which could serve as an interesting avenue for future research.

With the extension of our originally proposed simulation-based framework to support non-parametric prior distributions, we offer an elicitation framework, implemented in \emph{elicito} \citep{Bockting_elicito}, which can accommodate various types of elicitation methods within a single, unified structure. This facilitates the development of a standardized set of diagnostics and analysis tools, addressing a current gap in the literature \citep{mikkola2023prior}. 

In addition to accommodating different types of prior distributions, another strength of the framework is that \emph{expert knowledge} can be specified in various ways. This ranges from querying model parameters to observable quantities or any derived statistics, allowing the queried information to be tailored to the specific problem and the expertise of the domain expert. 
This design choice provides users with maximum flexibility in specifying the set of quantities to be elicited from the domain expert. However, it also places the responsibility on the user to define a valid and appropriate set of quantities to query from the expert \citep{simpson2017penalising}. We would argue that this flexibility is in general a desirable property of a method, as it avoids imposing specific modeling choices on the user. Thus, instead of requiring that the problem is framed in a predefined way, the method allows users to adapt it to their specific needs \citep{regenwetter2019tutorial,simpson2017penalising, scott2017prior}. 
However, in practice, this flexibility is only effective if good recommendations, guidelines, and diagnostics are available to help users define an appropriate set of quantities to query from the expert \citep{garthwaite2013prior}. 

A significant challenge in this context is assessing whether the chosen set of statistics is \emph{sufficiently informative} for the corresponding model. The goal is not necessarily to identify a unique prior that fully defines the generative model. Instead, the aim is to determine a set of equivalently plausible priors, supporting the assumption that a domain expert is unlikely to have a single ``true'' prior in mind \citep{winkler1967assessment}. This introduces at least two additional requirements for an elicitation method. First, it should include an efficient implementation that allows the training algorithm to be executed multiple times with different seeds, enabling the assessment of variability in the learned prior distributions. Second, the elicitation method should ideally offer additional diagnostics or advanced techniques, such as model selection or model averaging, to assist the user \citep{falconer2022methods}. In this work, we propose a model averaging method in which the weights calculated for averaging can also serve as diagnostics, helping to identify potential outliers within the set of learned priors. Identifying these edge cases could provide a valuable starting point for evaluating the plausibility of the learned priors in collaboration with a domain expert.

If the set of learned prior distributions includes degenerate cases, this signals the need for additional constraints. These constraints could involve eliciting further or alternative quantities (i.e., elicited statistics) from the expert and/or adding a regularization term to the total loss function to better guide the learning process.
To evaluate the informativeness of the elicited statistics, we use a sensitivity analysis as a preliminary evaluation tool \citep{o2004probability}. Additionally, we advocate for employing an oracle (i.e., simulating data from a known ground truth) and conducting multiple training replications with varying random seeds. These strategies provide a solid initial check to determine if the learned priors require further constraints. Overall, this discussion highlights the importance of developing comprehensive tutorials and resources to guide users effectively through the elicitation workflow as a key task for future work.

Other important design choices within our framework that require appropriate default settings include the specification of the discrepancy measure and the selection of the optimization approach, among other aspects. So far, in all simulation studies, we have used the maximum mean discrepancy \citep[MMD,][]{gretton2006kernel} to evaluate the discrepancy between the model-implied and expert-elicited statistics. The only exception is the correlation information, for which we employed an L2 loss.
One key advantage of the MMD is its flexibility in handling a wide range of statistics, from simple point estimates to complex histogram data. However, a notable limitation of the MMD is its high (i.e., quadratic) computational cost. In contrast, the L2 loss is computationally efficient but lacks the capability to process histogram data. Despite the higher computational cost, the MMD consistently showed very good performance and robustness across all simulation studies so far. Therefore, it appears to be a solid choice as a default discrepancy measure.
Future work should study the performance of different loss functions depending on different elicited statistics, with the goal to develop user guidelines for selecting an appropriate loss function.

Furthermore, in all simulation studies conducted so far, we used mini-batch SGD to optimize the prior network parameters. Given the generally good performance observed, we believe that this optimization method is a solid default choice. 
However, we aim to explore additional optimization approaches within our framework, such as Bayesian optimization (as employed by \citet{manderson2023translating}). Comparing these approaches in terms of performance and the level of hyperparameter tuning required could offer valuable insights into the conditions under which optimization method might be more advantageous than the other.

Finally, while our current methodology learns prior distributions based on expert data, it does currently not account for additional `meta' uncertainty that arises during the elicitation process. This includes uncertainties introduced by the elicitation technique itself or the expert's uncertainty regarding the `true' quantity \citep[see e.g.,][]{falconer2022methods, falconer2024eliciting, stefan2022practical}. We acknowledge that these are important sources of information that should be incorporated in future developments of our method. One promising approach is to view this problem from a Bayesian perspective and treat the proposed workflow as a Bayesian inference problem. This idea was already proposed by \citet{mikkola2023prior} and referred to as the \emph{supra-Bayesian approach}. We believe this is an important and valuable direction for the next steps in our elicitation method.

\newpage

\section*{Declarations}
\paragraph{Acknowledgements}
We thank Dmytro Perepolkin and Luna Fazio for their valuable comments and suggestions that greatly improved the manuscript.
\paragraph{Funding}
Not applicable.
\paragraph{Competing interests}
The authors have no competing interests to declare that are relevant to the content of this article.
\paragraph{Consent to participate}
Not applicable.
\paragraph{Consent for publication}
Not applicable.
\paragraph{Data availability}
The simulation results are available on OSF at \url{https://osf.io/xrzh6/}.
\paragraph{Materials availability}
Supplementary material is available on GitHub at \url{https://github.com/florence-bockting/non-parametric-prior}.
\paragraph{Code availability}
Code for running the simulations and plotting the results is available on GitHub at \url{https://github.com/florence-bockting/non-parametric-prior}. Simulations and most plots are based on/provided by our Python package \emph{elicito} \citep[][v0.3.1]{Bockting_elicito}. To ensure the reproducibility of our results, the specific version of \emph{elicito} used in this study has been archived on Zenodo and is accessible via the following DOI \url{https://doi.org/10.5281/zenodo.15241853}.
\paragraph{Author contribution}
Conceptualization: PB, SR, FB, Methodology: PB, FB, Software: FB, Writing (Original Draft): FB, Writing (Editing): FB, SR, PB, Writing (Review): FB, PB, Visualization: FB, Supervision: PB, Funding acquisition: PB

\newpage
\bibliography{manuscript_submission/bibliography.bib}

\begin{appendices}
\newpage
\section{Simulation Studies}\label{secA2}
\subsection{Training time}\label{subsecA1}
The median training time across the 30 replications ($\pm$ sd of training time):
\begin{itemize}
    \item \textbf{Binomial model}: 6.87 minutes ($\pm$ 1.45).
    \item \textbf{Independent normal scenario}: 10.18 minutes ($\pm$ 1.68).
    \item \textbf{Skewed normal scenario}: 14.45 minutes ($\pm$ 2.80).
    \item \textbf{Correlated normal scenario}: 10.10 minutes ($\pm$ 0.17).
\end{itemize}

\begin{figure}[ht]
\subsection{Independent normal scenario}\label{subsecA2}
    \centering
    \includegraphics[width=0.9\linewidth] {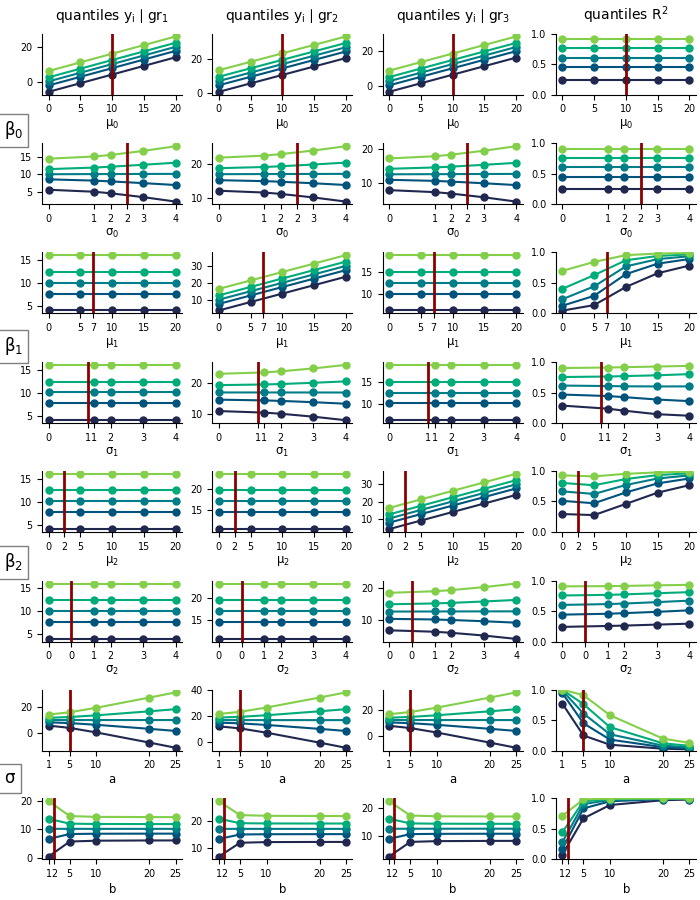}
    \caption{\emph{Normal model / Scenario 1 (M2) --- Sensitivity Analysis}: Rows represent hyperparameters of each model parameter. Columns represent elicited statistics: five quantiles for $y \mid \text{gr}_i$ with $i=1,2,3$ and for $R^2$. Quantiles are depicted in different colors. In each row, the corresponding hyperparameter, is varied across the range shown on the x-axis, while all other hyperparameters are held constant at their true values, indicated by the red vertical line.}
    \label{fig:sensitivity-independent-normal}
\end{figure}
\begin{figure}[ht]
    \centering
    \includegraphics[width=0.8\linewidth] {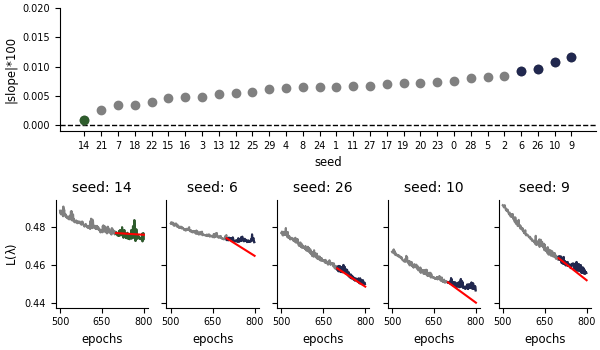}
    \caption{\emph{Normal model / Scenario 1 (M2) --- Preliminary Convergence Check}: Upper plot displays the absolute slope of the total loss trajectory (scaled by a factor of 100) over the last 100 epochs on the y-axis, with each replication shown along the x-axis. The replications (i.e., seeds) are arranged in ascending order based on the magnitude of the absolute slope. The best-performing model is seed 14, while the worst-performing models are seeds 6, 26, 10, and 9. The lower plot illustrates the loss trajectories of the best- and worst-performing seeds over the last 300 epochs. The loss updates for the final 100 epochs are highlighted in green (best model) and blue (worst models), with a red line segment indicating the respective slope.}
    \label{fig:convergence-1-independent-normal}
    \vspace{0.5cm}
    \includegraphics[width=0.7\linewidth] {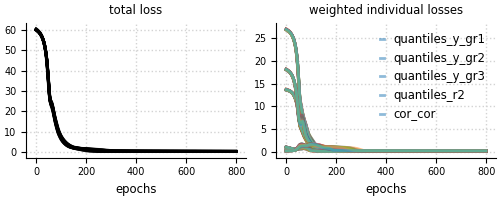}
    \caption{\emph{Normal model / Scenario 1 (M2) --- Visual convergence check 1}: Updating trajectory across epochs for the total loss (left) and the individual (weighted) loss components (right).}
    \label{fig:convergence-2-independent-normal}
\end{figure}

\begin{figure}[ht]
    \centering
    \includegraphics[width=0.7\linewidth] {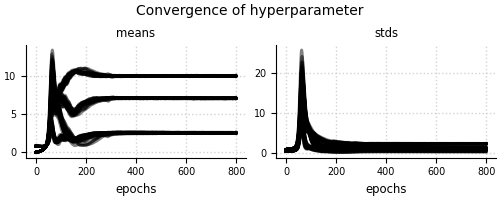}
    \caption{\emph{Normal model / Scenario 1 (M2) --- Visual convergence check 2}: Updating trajectory across epochs for the mean (left) and standard deviation (right) of the final learned marginal priors for all 30 replications.}
    \label{fig:convergence-3-independent-normal}
\end{figure}
\begin{figure}[ht]
    \centering
    \includegraphics[width=0.8\linewidth] {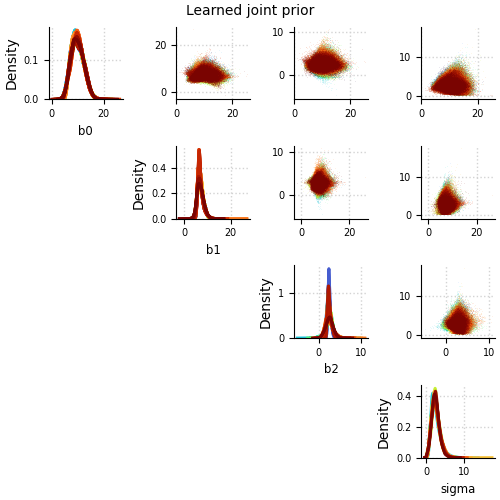}
    \caption{\emph{Normal model / Scenario 1 (M2) --- Joint prior}: Each replication is visualized in a different color.}
    \label{fig:joint-prior-independent-normal}
\end{figure}

\begin{figure}[ht]
\subsection{Skewed normal scenario}\label{subsecA3}
    \centering
    \includegraphics[width=0.8\linewidth] {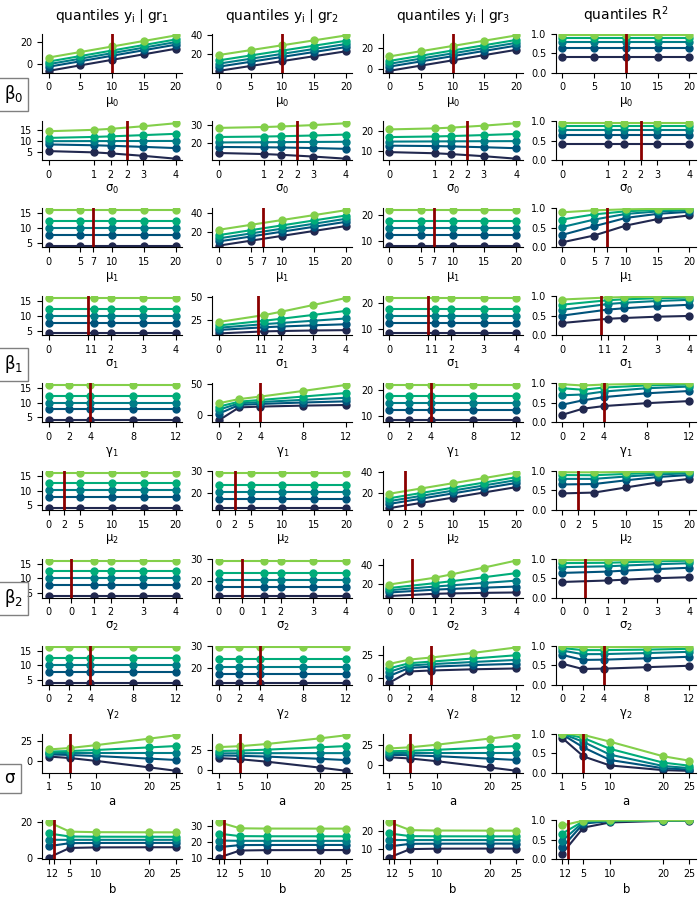}
    \caption{\emph{Normal model / Scenario 2 (M3) --- Sensitivity Analysis}: Rows represent hyperparameters of each model parameter. Columns represent elicited statistics: five quantiles for $y \mid \text{gr}_i$ with $i=1,2,3$ and for $R^2$. Quantiles are depicted in different colors. In each row, the corresponding hyperparameter, is varied across the range shown on the x-axis, while all other hyperparameters are held constant at their true values, indicated by the red vertical line. }
    \label{fig:sensitivity-skewed-normal}
\end{figure}

\begin{figure}[ht]
    \centering
    \includegraphics[width=0.8\linewidth] {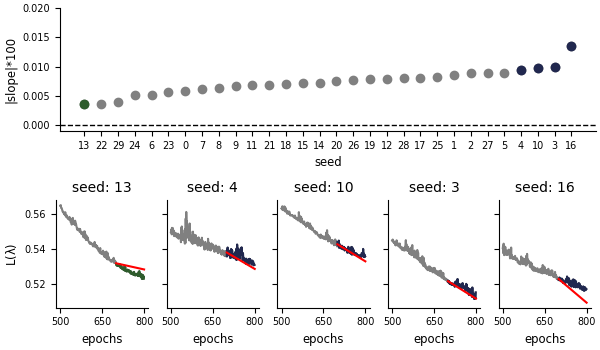}
    \caption{\emph{Normal model / Scenario 2 (M3) --- Preliminary Convergence Check for all 30 replications}: The upper plot displays the absolute slope of the total loss trajectory (scaled by a factor of 100) over the last 100 epochs on the y-axis, with each replication shown along the x-axis. The replications (i.e., seeds) are arranged in ascending order based on the magnitude of the absolute slope. The best-performing model is seed 13, while the four worst-performing models are seeds 4, 10, 3, and 16. The lower plot illustrates the loss trajectories of the best- and worst-performing seeds over the last 300 epochs. The loss updates for the final 100 epochs are highlighted in green (best seed) or blue (worst seeds). The red line segment indicates the respective slope.}
    \label{fig:convergence-1-skewed-normal}
    \vspace{0.5cm}
    \includegraphics[width=0.7\linewidth] {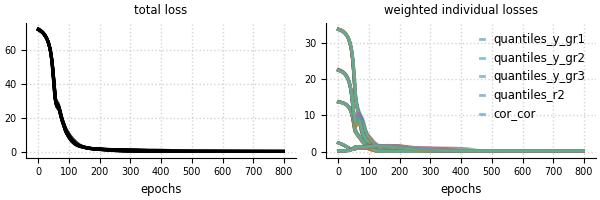}
    \caption{\emph{Normal model / Scenario 2 (M3) --- Visual convergence check 1}: Updating trajectory across epochs for the total loss (left) and the individual (weighted) loss components (right).}
    \label{fig:convergence-2-skewed-normal}
\end{figure}

\begin{figure}[ht]
    \centering
    \includegraphics[width=0.7\linewidth] {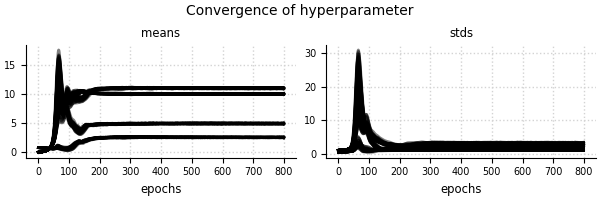}
    \caption{\emph{Normal model / Scenario 2 (M3) --- Visual convergence check 2}: Updating trajectory across epochs for the mean (left) and standard deviation (right) of the marginal priors.}
    \label{fig:convergence-3-skewed-normal}
\end{figure}

\begin{figure}[ht]
    \centering
    \includegraphics[width=0.8\linewidth] {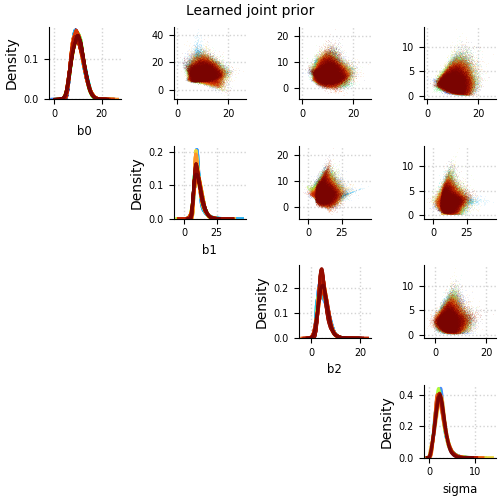}
    \caption{\emph{Normal model / Scenario 2 (M3) --- Visual convergence check 2}: Joint prior distribution. Each replication is visualized in a different color.}
    \label{fig:joint-prior-skewed-normal}
\end{figure}

\begin{figure}[ht]
\subsection{Correlated normal scenario}\label{subsecA4}
    \centering
    \includegraphics[width=0.8\linewidth] {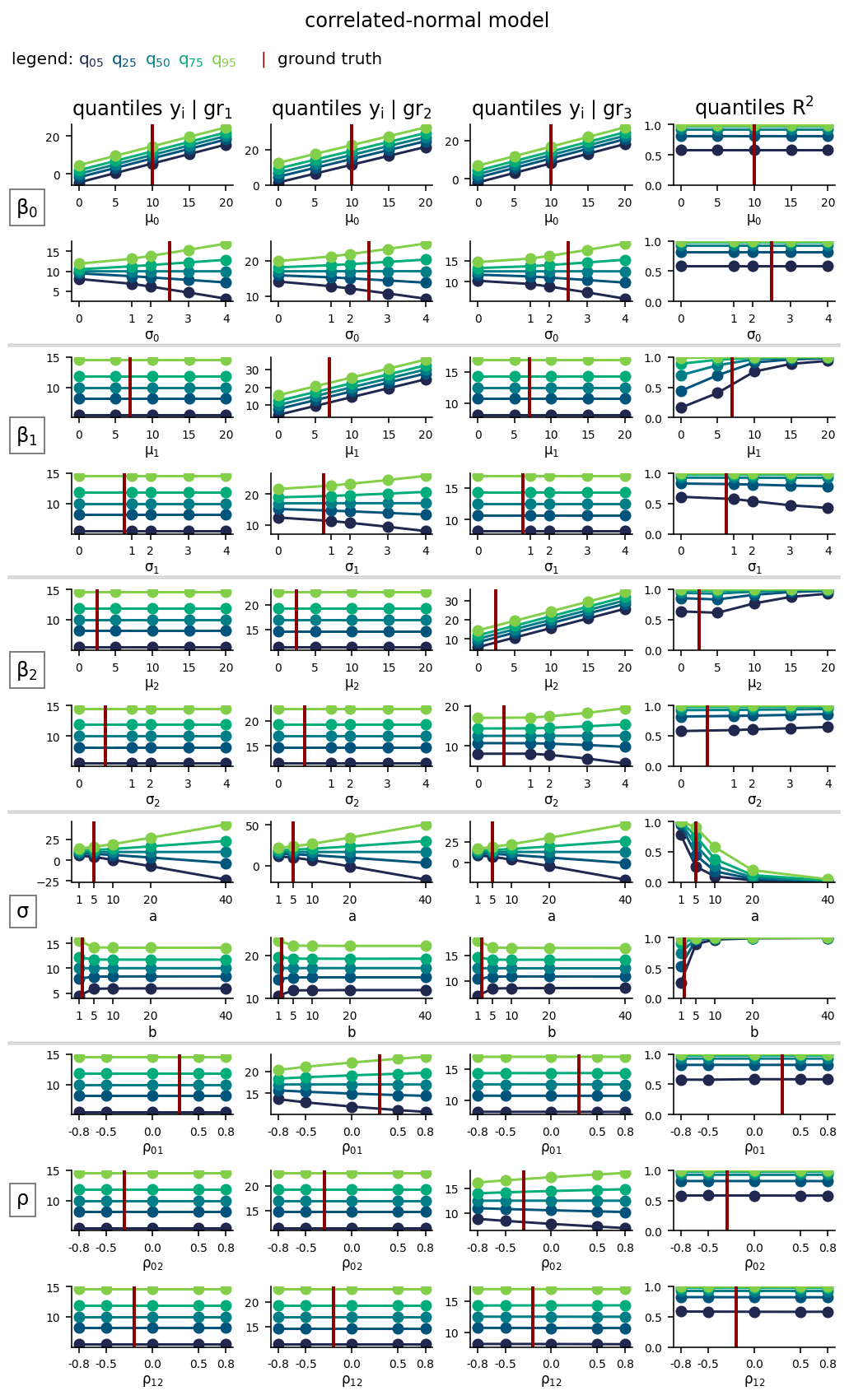}
    \caption{\emph{Normal model / Scenario 3 (M4) --- Sensitivity Analysis}: Rows represent hyperparameters of each model parameter. Columns represent elicited statistics: five quantiles for $y \mid \text{gr}_i$ with $i=1,2,3$ and for $R^2$. Quantiles are depicted in different colors. In each row, the corresponding hyperparameter, is varied across the range shown on the x-axis, while all other hyperparameters are held constant at their true values, indicated by the red vertical line.}
    \label{fig:sensitivity-correlated-normal}
\end{figure}

\begin{figure}[ht]
    \centering
    \includegraphics[width=0.8\linewidth] {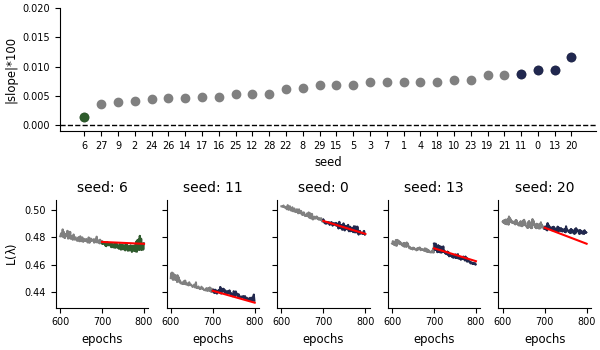}
    \caption{\emph{Normal model / Scenario 3 (M4) --- Preliminary Convergence Check for all 30 replications}: The upper plot displays the absolute slope of the total loss trajectory (scaled by a factor of 100) over the last 100 epochs on the y-axis, with each replication shown along the x-axis. The replications (i.e., seeds) are arranged in ascending order based on the magnitude of the absolute slope. The best-performing model is seed 6, while the worst-performing models are seeds 11, 0, 13, and 20. The lower plot illustrates the loss trajectories of the best- and worst-performing seeds over the last 200 epochs. The loss updates for the final 100 epochs are highlighted in green for the best model and in blue for the worst models, with a red line segment indicating the respective slope.}
    \label{fig:convergence-1-correlated-normal}
    \vspace{0.5cm}
    \includegraphics[width=0.7\linewidth] {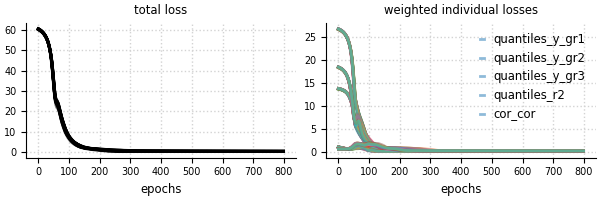}
    \caption{\emph{Normal model / Scenario 3 (M4) --- Visual convergence check 1}: Updating trajectory across epochs for the total loss (left) and the individual (weighted) loss components (right).}
    \label{fig:convergence-2-correlated-normal}
\end{figure}

\begin{figure}[ht]
    \centering
    \includegraphics[width=0.7\linewidth] {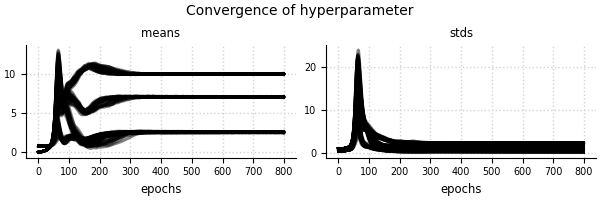}
    \caption{\emph{Normal model / Scenario 3 (M4) --- Visual convergence check 1}: Updating trajectory across epochs for the mean (left) and standard deviation (right) of the marginal priors.}
    \label{fig:convergence-3-correlated-normal}
\end{figure}

\begin{figure}[ht]
    \centering
    \includegraphics[width=0.8\linewidth] {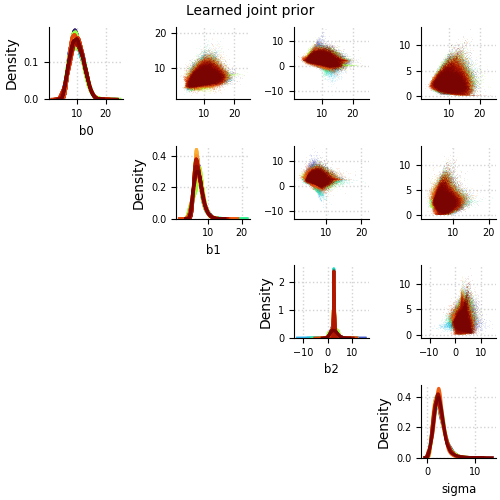}
    \caption{\emph{Normal model / Scenario 3 (M4) --- Visual convergence check 2}: Joint prior distribution. Each replication is visualized in a different color.}
    \label{fig:joint-prior-correlated-normal}
\end{figure}

\end{appendices}

\end{document}